\documentclass[prc,preprint,showpacs,showkeys,lengthcheck,
  nofootinbib,tightenlines,onecolumn,notitlepage,preprintnumbers,superscriptaddress]{revtex4-1}

\usepackage[utf8]{inputenc}
\usepackage{newtxtext,newtxmath}
\usepackage[mathcal]{euscript}

\usepackage{graphicx}
\usepackage[usenames,dvipsnames]{xcolor}
\graphicspath{{figs/}}
\usepackage{cancel}

\usepackage{hyperref}
\hypersetup{colorlinks=true,citecolor=blue,linkcolor=blue,urlcolor=blue}

\usepackage{diagbox}

\usepackage{amsmath,amssymb,amsfonts,dsfont}
\usepackage{slashed}
\usepackage{bm}
\usepackage{array}

\newcommand{\blue}[1]{{#1}}

\newcommand{\tabincell}[2]{\begin{tabular}{@{}#1@{}}#2\end{tabular}}
\begin{document}
\title{Covariant multipole expansion of local currents for massive states of any spin }

\author{Sabrina Cotogno}

\email{sabrina.cotogno@polytechnique.edu}
\affiliation{CPHT, CNRS, Ecole Polytechnique, Institut Polytechnique de Paris, Route de Saclay, 91128 Palaiseau, France}

\author{C\'edric Lorc\'e}
\email{cedric.lorce@polytechnique.edu}
\affiliation{CPHT, CNRS, Ecole Polytechnique, Institut Polytechnique de Paris, Route de Saclay, 91128 Palaiseau, France}
	
	\author{Peter Lowdon}
\email{peter.lowdon@polytechnique.edu}
\affiliation{CPHT, CNRS, Ecole Polytechnique, Institut Polytechnique de Paris, Route de Saclay, 91128 Palaiseau, France}

	\author{Manuel Morales}
\email{malvarado@student.ethz.ch}
\affiliation{ETH Z\"urich, R\"amistrasse 101, 8092 Z\"urich, Switzerland}

\begin{abstract}

We study the structure of scalar, vector, and tensor currents for on-shell massive particles of any spin. When considering higher values for the spin of the particle, the number of form factors (FFs) involved in the decomposition of the matrix elements associated with these local currents increases. We identify all the fundamental structures that give rise to the independent FFs, systematically for any spin value. These structures can be conveniently organised using an expansion in  covariant multipoles, built solely from the Lorentz generators. This approach allows one to uniquely identify the terms which are universal and those that arise because of spin.
We derive counting rules which relate the number of FFs to the total spin $j$ of the state, showing explicitly that these rules match all the well-known cases up to spin 2.
\end{abstract}

\maketitle
\onecolumngrid

\section{Introduction}
\label{sec:intro}

Matrix elements representing the interaction of quantum states with local currents are parametrised in terms of form factors (FFs). The most known examples are the electromagnetic and gravitational FFs, which are related to the electromagnetic current and the energy-momentum tensor (EMT), respectively, and which carry important information on the nature of the system. In the particular case of QCD, these FFs are fundamental observables which contain a rich information about the internal structure of hadrons, ranging from their electromagnetic properties to the spatial and angular momentum distributions of their internal constituents~\cite{Polyakov:2002yz,Leader:2013jra,Lorce:2018zpf,Lorce:2017wkb,Lorce:2018egm,Polyakov:2018zvc}.

In hadronic physics, a large amount of attention has been historically given to the proton, due to its abundance as a stable particle and its central role in the building of visible matter. However, recently there has been an increasing interest in the study of higher-spin hadrons, as unique tools to study the dynamics of internal constituents beyond the degrees of freedom typical of a single spin-$\frac{1}{2}$ nucleon (proton and neutron)~\cite{Jaffe:1989xy,Efremov:1981vs,Efremov:1994xf,Holstein:2006ud,Abidin:2008ku,Alexandrou:2008bn,Alexandrou:2009hs,Taneja:2011sy,Boeglin:2015cha,Boer:2016xqr,Kumano:2016ude,Detmold:2016gpy,Winter:2017bfs,Detmold:2019ghl,Cosyn:2019aio,Polyakov:2019lbq}.  
Although measurements of the FFs for higher-spin particles would be experimentally challenging and hardly feasible in a foreseeable future, investigating higher-spin problems nevertheless remains desirable from a broader theoretical point of view~\cite{Cotogno:2019xcl,Lorce:2019sbq,Bautista:2019tdr}. For instance, old-standing problems concerning the fundamental interactions for particles of arbitrary spin  have been explored in the past, with different approaches and techniques, but they still lack a global theoretical description. 
Only in specific cases, such as the scalar and electromagnetic interaction, has the formalism developed to study massive spin-$\frac{1}{2}$ particles been extended to higher spins: the precise rule which links the number of electromagnetic FFs with the value $j$ of the spin was established long ago and unanimously accepted in the literature~\cite{Cheshkov1963,Scadron:1968zz,Williams:1970ms,Lorce:2009bs}.  
However, a similar systematic and unambiguous counting of FFs is missing for the tensor currents of higher rank (in particular for the EMT), and a few past attempts have led to various answers, see e.g.~\cite{Cheshkov1963,Scadron:1968zz,Williams:1970ms}

Finding the most general expression for the EMT, which is not relegated to a spin-by-spin analysis, is extremely useful to shed light on the universal properties of particles.
Recently, it was rigorously proven~\cite{Cotogno:2019xcl,Lorce:2019sbq} that constraints on the gravitational FFs in the limit of zero momentum transfer, historically associated to spin-$\frac{1}{2}$ particles, are actually universal for states of arbitrary spin. These constraints are solely related to Poincar\'e symmetry, and hold independently of the spin of the particles and their mass. 
The crucial starting point to this proof is the realisation that, for all physical states, the conserved (truncated) EMT matrix element can be expressed as the sum of a spin-independent part and a term linear in the Lorentz generators in the given representation~\cite{Boulware:1974sr}. The expansion truncates to terms at most linear in the momentum transfer $\Delta$, because they are the only ones constrained by the generators of Poincar\'e symmetry~\cite{Lowdon:2017idv,Cotogno:2019xcl,Lorce:2019sbq}. A complete expression for the EMT would also include terms which depend on higher powers of $\Delta$ and the Lorentz generators. 

A natural question that arises is whether it is possible to characterise the role of the spin of the state in shaping the structure of the EMT and, consequently, the number of gravitational FFs, analogously to the vector current case. In other words, one might wonder what is the best systematic approach to find the complete parametrisation for a given operator (scalar, vector, and tensor) such that the FF counting depends only on the total spin $j$. Additionally, one might wonder how the expansion changes with the rank of the operator, and which terms appear purely due to the spin of the particle. \newline

In this work we address this question and present the complete parametrisations for the matrix elements of scalar, vector, and rank-2 tensor currents for massive states of arbitrary spin. 
Following the existing literature on the vector current case~\cite{Lorce:2009bs}, we first derive in Section~\ref{sec:param1} a parametrisations using a tensor product approach. We single out all the possible ``core'' or ``seed'' structures, i.e.\ Lorentz structures that contribute to the expansion of the matrix element for a given type of local current, and associate to them a ``tower'' of elements, whose number depends on the spin of the particle. 
This approach leads to the explicit expression for the EMT parametrisation in a given spin representation, and enables one to determine the number of FFs as a function of the spin. One limitation though is that the seeds are specific to each operator (scalar, vector, and tensor), and their linear independence needs to be checked explicitly in order to avoid incomplete or overcomplete expressions.

The choice of basis is of course arbitrary, and different parametrisations are related to each other and must provide the same counting rule.
In the spirit of Ref.~\cite{Cotogno:2019xcl} and with the aim of looking for the most general way to find the EMT parametrisation we present an alternative approach based on the covariant multipole expansion developed in Sections~\ref{sec:multipole} and~\ref{sec:covariantparam}.  
One can take advantage of the fact that all physical observables are elements of the Lorentz group. A natural basis for the parametrisation is therefore based on covariant multipoles, built from symmetric and traceless products of Lorentz generators in a given spin representation. They are the covariant extension of the non-relativistic multipoles of the $\mathfrak{su}(2)$ Lie algebra, built from the generators of rotations. A remarkable advantage of the multipole expansion is that it truncates at some given order. In particular, only the first $2j+1$ multipoles are non-zero, whereas higher multipoles vanish. In addition, each new multipole is guaranteed to be independent of the previous ones. Being formed by symmetrised products of Lorentz generators, the only non-vanishing $2j+1$ multipoles are operators with given symmetry properties on each pair of Lorentz indices.  Starting from this basis of linearly independent multipoles, which is common to all operators (scalar, vector, tensor) for a given state of spin $j$, we build the coefficients of the expansion depending on the symmetry properties of the problem and on the relevant operator. This procedure leads to a systematic counting of FFs. Interestingly, we can show that the counting changes in a non-trivial way when going from lower to higher-rank operators. 

The strength of the multipole expansion lies in its generality and conceptual intuitiveness. However,
the two approaches developed in this paper mutually aid each other in reaching the final counting rule.  
As a natural follow-up of this work, one can aspire to extend the counting to higher-rank operators~\cite{Hagler:2004yt} and, interestingly, to non-local currents such as those entering parton distributions like PDFs, GPDs, and TMDs used to describe observables in hadronic physics. 

Finally, we also include two appendices. In the first one we describe the explicit construction of higher-spin polarisation tensors. In the second one we derive a large set of exact and on-shell identities, which are used to eliminate redundant Lorentz structures in the parametrisations.

\section{Parametrisation using the tensor product approach}\label{sec:param1}

Matrix elements representing a generic local rank-$k$ current for arbitrary spin states of mass $M$ can be written as
\begin{equation}
\langle p',\lambda'|\hat O^{\mu_1\cdots\mu_k}(0)|p,\lambda\rangle=\overline\eta(p',\lambda')\,O^{\mu_1\cdots\mu_k}(P,\Delta)\,\eta(p,\lambda).
\label{ElMatEl}
\end{equation}
For later convenience we introduce the average four-momentum $P=(p'+p)/2$ and the four-momentum transfer $\Delta=p'-p$ satisfying the on-shell conditions $P^2+\Delta^2/4=M^2$ and $P\cdot\Delta=0$. The polarisation of physical states is described by a generalised polarisation tensor (GPT) $\eta(p,\lambda)$ as in~\cite{Lowdon:2017idv,Cotogno:2019xcl,Lorce:2019sbq}. GPTs are defined such that the covariant density matrix in a given representation of the Lorentz group
\begin{equation}
\rho^A_{\phantom{A}B}(p,\lambda,\lambda')=\eta^A(p,\lambda) \,\overline\eta_B(p,\lambda')
\end{equation}
has normalisation $\text{Tr}[\rho(p,\lambda,\lambda')]=\delta_{\lambda\lambda'}$. An irreducible representation of the Lorentz group $(j_1,j_2)$ is in general reducible under the subgroup of rotations $SU(2)$. It involves all spin values $j$ obtained by the standard composition rule of angular momenta
\begin{equation}
j_1\otimes j_2=\bigoplus_{j=|j_1-j_2|}^{j_1+j_2}j.
\end{equation}
A physical particle can therefore be described by some Lorentz representation $(j_1,j_2)$ provided that subsidiary conditions are imposed to get rid of the unwanted, unphysical spin representations. 

Let us consider a particle of mass $M$ and spin $j$. When $j=n$ is integer, we choose to work with the $(\frac{n}{2},\frac{n}{2})$ representation where the GPT  $\eta(p,\lambda)\sim\varepsilon_{\alpha_1\cdots\alpha_n}(p,\lambda)$ is totally symmetric, traceless and satisfies the subsidiary condition
\begin{equation}\label{subcond1}
p^\alpha \varepsilon_{\alpha\alpha_2\cdots\alpha_n}(p,\lambda)=0.
\end{equation}
When $j=n+\frac{1}{2}$ is half-integer, we choose to work with the $(\frac{n+1}{2},\frac{n}{2})\oplus(\frac{n}{2},\frac{n+1}{2})$ representation where the GPT $\eta(p,\lambda)\sim u_{\alpha_1\cdots\alpha_n}(p,\lambda)$ is totally symmetric, traceless and satisfies
the subsidiary conditions\footnote{Note that the first condition is superfluous since it can be derived from the other two.}
\begin{equation}\label{subcond2}
\begin{aligned}
p^\alpha u_{\alpha\alpha_2\cdots\alpha_n}(p,\lambda)&=0,\\
\left(p\!\!\!\slash\,-M\right)u_{\alpha_1\cdots\alpha_n}(p,\lambda)&=0,\\
\gamma^\alpha u_{\alpha\alpha_2\cdots\alpha_n}(p,\lambda)&=0.
\end{aligned}
\end{equation}
The subsidiary conditions~\eqref{subcond1} and~\eqref{subcond2} simply ensure that the number of degrees of freedom is $2j+1$. For more details on the construction of these GPTs, see Appendix~\ref{AppA}. 
\newline

When $j=n$ is integer, the expression~\eqref{ElMatEl} reads more explicitly in the $(\frac{n}{2},\frac{n}{2})$ representation
\begin{equation}
\langle p',\lambda'|\hat O^{\mu_1\cdots\mu_k}(0)|p,\lambda\rangle=(-1)^n\,\varepsilon^*_{\alpha'_1\cdots\alpha'_n}(p',\lambda')\,O^{\mu_1\cdots\mu_k,\alpha'_1\cdots\alpha'_n\alpha_1\cdots\alpha_n}(P,\Delta)\,\varepsilon_{\alpha_1\cdots\alpha_n}(p,\lambda).
\end{equation}
The overall $(-1)^n$ factor ensures that GPTs are properly normalised $\bar\eta_A(p,\lambda)\eta^A(p,\lambda)=(-1)^n\,\varepsilon^*_{\alpha_1\cdots\alpha_n}(p,\lambda)\varepsilon^{\alpha_1\cdots\alpha_n}(p,\lambda)=1$. Thanks to the Lorentz invariance of the theory, the tensor $O^{\mu_1\cdots\mu_k,\alpha'_1\cdots\alpha'_n\alpha_1\cdots\alpha_n}$ can be expressed as a sum of Lorentz tensors built out of the Minkowski metric $g_{\mu\nu}$, the totally antisymmetric Levi-Civita pseudo-tensor\footnote{We use the convention $\epsilon_{0123}=+1$.} $\epsilon_{\mu\nu\rho\sigma}$, and the four-vectors of the problem $P^\mu$ and $\Delta^\mu$. Each of these Lorentz structures is multiplied by a Lorentz scalar function of $t=\Delta^2$, and are referred to as form factors (FFs).

When $j=n+\frac{1}{2}$ is half-integer, the expression~\eqref{ElMatEl} can in a similar way be written more explicitly in the $(\frac{n+1}{2},\frac{n}{2})\oplus(\frac{n}{2},\frac{n+1}{2})$ representation as
\begin{equation}
\langle p',\lambda'|\hat O^{\mu_1\cdots\mu_k}(0)|p,\lambda\rangle=(-1)^n\,\overline u_{\alpha'_1\cdots\alpha'_n}(p',\lambda')\,O^{\mu_1\cdots\mu_k,\alpha'_1\cdots\alpha'_n\alpha_1\cdots\alpha_n}(P,\Delta)\,u_{\alpha_1\cdots\alpha_n}(p,\lambda).
\end{equation}
The difference with the integer-spin case is that the tensor $O^{\mu_1\cdots\mu_k,\alpha'_1\cdots\alpha'_n\alpha_1\cdots\alpha_n}$ is now a matrix in Dirac space\footnote{Dirac indices are omitted for better legibility.}. We can therefore also use the Dirac matrices $\gamma^\mu$ and their products to construct tensor structures. The identification of a proper basis of structures is consequently even more complex.

Discrete symmetries constrain further the operators~\cite{Meissner:2009ww}. Hermiticity requires operators carrying some Lorentz indices $\hat O^{\mu_1\cdots\mu_k}(x)$ to satisfy
\begin{equation}
\langle p',\lambda'|\hat O^{\mu_1\cdots\mu_k}(x)|p,\lambda\rangle=\langle p,\lambda|\hat O^{\mu_1\cdots\mu_k}(x)|p',\lambda'\rangle^*.
\end{equation}
We will also impose \textsf{P} and \textsf{T} symmetries, and restrict ourselves to operators with positive intrinsic parity and time-reversal properties. These symmetries imply the following constraints
\begin{equation}\label{dissymm1}
\begin{aligned}
O^{\mu_1\cdots\mu_k,\alpha'_1\cdots\alpha'_n\alpha_1\cdots\alpha_n}(P,\Delta)&=[O^{\mu_1\cdots\mu_k,\alpha_1\cdots\alpha_n\alpha'_1\cdots\alpha'_n}(P,-\Delta)]^*\\
&=O^{\bar\mu_1\cdots\bar\mu_k,\bar\alpha'_1\cdots\bar\alpha'_n\bar\alpha_1\cdots\bar\alpha_n}(\bar P,\bar\Delta)\\
&=[O^{\bar\mu_1\cdots\bar\mu_k,\bar\alpha'_1\cdots\bar\alpha'_n\bar\alpha_1\cdots\bar\alpha_n}(\bar P,\bar\Delta)]^*,    
\end{aligned}
\end{equation}
when $j=n$ is integer, and
\begin{equation}\label{dissymm2}
\begin{aligned}
O^{\mu_1\cdots\mu_k,\alpha'_1\cdots\alpha'_n\alpha_1\cdots\alpha_n}(P,\Delta)&=\gamma^0 [O^{\mu_1\cdots\mu_k,\alpha_1\cdots\alpha_n\alpha'_1\cdots\alpha'_n}(P,-\Delta)]^\dag\gamma^0\\
&=\gamma^0 O^{\bar\mu_1\cdots\bar\mu_k,\bar\alpha'_1\cdots\bar\alpha'_n\bar\alpha_1\cdots\bar\alpha_n}(\bar P,\bar\Delta)\gamma^0\\
&=(i\gamma^1\gamma^3)[O^{\bar\mu_1\cdots\bar\mu_k,\bar\alpha'_1\cdots\bar\alpha'_n\bar\alpha_1\cdots\bar\alpha_n}(\bar P,\bar\Delta)]^*(i\gamma^1\gamma^3),    
\end{aligned}
\end{equation}
when $j=n+\frac{1}{2}$ is half-integer\footnote{Here Dirac spinors are chosen in the standard or Dirac representation.}. In these expressions, we used the convenient notation $\bar a^\mu=a^{\bar\mu}=(a^0,-\vec a)$. Factors of $i$ appearing in the tensor structures are chosen so that FFs are real-valued functions. Because of the symmetry, tracelessness and subsidiary conditions satisfied by the GPTs, not all of the possible tensor structures are independent. We therefore have to carefully identify a linearly independent subset. A list of identities used to obtain our parametrisations is presented in Appendix~\ref{AppB}.

\subsection{Scalar operator}

The simplest operator is the scalar $\hat N(x)$. A typical example is the condensate operator $\hat N_q(x)=\hat{\overline\psi}(x)\hat\psi(x)$. When $j=n$ is integer, we find that the elastic matrix elements can be written in terms of the following basis
\begin{equation}\label{scalarop}
N^{\alpha'_1\cdots\alpha'_n\alpha_1\cdots\alpha_n}(P,\Delta)={\blue{2M}}\sum_{(k,n)}F^S_k(t),
\end{equation}
where the strange sum stands for
\begin{equation}
\sum_{(k,n)}\equiv\sum_{k=0}^n\left[\prod_{i=1}^k\left(-\frac{\Delta^{\alpha'_i}\Delta^{\alpha_i}}{\blue{2}M^2}\right)\prod_{i=k+1}^n g^{\alpha'_i\alpha_i}\right],
\label{strange_sum}
\end{equation}
and the mass factor accounts for the correct mass dimension of the operator, since we have normalised the GPTs to 1.
The same basis can be used to write the elastic matrix elements when $j=n+\frac{1}{2}$ is half-integer. The number of FFs associated with a scalar operator is therefore $n+1$, where $n=\lfloor j\rfloor$ is the floor of the spin, i.e. the largest integer  smaller or equal to $j$.

For spin-$0$, one has:
\begin{equation}
\langle p',\lambda'|\hat N(0)|p,\lambda\rangle={\blue{2M}} F^S_0(t).
\end{equation}

For spin\footnote{\blue{For convenience, we choose to normalise the Dirac spinors as $\overline u(p,\lambda')u(p,\lambda)=1$ instead of $\overline u(p,\lambda')u(p,\lambda)=2M$.}}-$\frac{1}{2}$:
\begin{equation}
\langle p',\lambda'|\hat N(0)|p,\lambda\rangle={\blue{2M}}\,\overline u(p',\lambda')u(p,\lambda)\,F^S_0(t).
\end{equation}

For spin-$1$:
\begin{equation}
\langle p',\lambda'|\hat N(0)|p,\lambda\rangle=-{\blue 2M}\,\varepsilon^*_{\alpha'}(p',\lambda')\left[g^{\alpha'\alpha}\,F^S_0(t)-\frac{\Delta^{\alpha'}\Delta^\alpha}{\blue 2M^2}\,F^S_1(t)\right]\varepsilon_{\alpha}(p,\lambda).
\end{equation}

For spin-$\frac{3}{2}$:
\begin{equation}
\langle p',\lambda'|\hat N(0)|p,\lambda\rangle=-{\blue 2M}\,\overline u_{\alpha'}(p',\lambda')\left[g^{\alpha'\alpha}\,F^S_0(t)-\frac{\Delta^{\alpha'}\Delta^\alpha}{\blue 2M^2}\,F^S_1(t)\right]u_{\alpha}(p,\lambda).
\end{equation}

For spin-$2$:
\begin{equation}
\langle p',\lambda'|\hat N(0)|p,\lambda\rangle={\blue 2M}\,\varepsilon^*_{\alpha'_1\alpha'_2}(p',\lambda')\left[g^{\alpha'_1\alpha_1}g^{\alpha'_2\alpha_2}\,F^S_0(t)-\frac{\Delta^{\alpha'_1}\Delta^{\alpha_1}}{\blue 2M^2}\,g^{\alpha'_2\alpha_2}\,F^S_1(t)+\frac{\Delta^{\alpha'_1}\Delta^{\alpha_1}}{\blue 2M^2}\,\frac{\Delta^{\alpha'_2}\Delta^{\alpha_2}}{\blue 2M^2}\,F^S_2(t)\right]\varepsilon_{\alpha_1\alpha_2}(p,\lambda).
\end{equation}

\subsection{Vector operator}
\label{Vector_tensapproach}

Let us now consider the (four-)vector operator $\hat J^\mu(x)$. A typical example is the charge current $\hat J^\mu_e(x)=e\hat{\overline\psi}(x)\gamma^\mu\hat\psi(x)$. Various parametrisations have been proposed in the literature for spin-$\frac{1}{2}$~\cite{Foldy:1952zz}, $1$~\cite{Arnold:1979cg}, $\frac{3}{2}$~\cite{Nozawa:1990gt,Pascalutsa:2006up} and higher~\cite{Cheshkov1963,Scadron:1968zz,Williams:1970ms,Lorce:2009bs}. We find that its elastic matrix elements can be written in terms of the following basis 
\begin{equation}\label{Jmuint}
\begin{aligned}
J^{\mu,\alpha'_1\cdots\alpha'_n\alpha_1\cdots\alpha_n}(P,\Delta)&={\blue 2}P^\mu\sum_{(k,n)}F^V_{1,k}(t)\\
&-\left(g^{\mu\alpha'_n}\Delta^{\alpha_n}-g^{\mu\alpha_n}\Delta^{\alpha'_n}\right)\sum_{(k,n-1)}F^V_{2,k}(t),
\end{aligned}
\end{equation}
when $j=n$ is integer, and
\begin{equation}\label{Jmuhalfint}
\begin{aligned}
J^{\mu,\alpha'_1\cdots\alpha'_n\alpha_1\cdots\alpha_n}(P,\Delta)&={\blue 2}P^\mu\sum_{(k,n)}F^V_{1,k}(t)\\
&+{\blue i}\sigma^{\mu\nu}\Delta_\nu\sum_{(k,n)}F^V_{2,k}(t)
\end{aligned}
\end{equation}
and when $j=n+\frac{1}{2}$ is half-integer.
The number of FFs associated with a vector operator is therefore $2j+1$, as already established long ago~\cite{Durand:1962zza,Scadron:1968zz,Fearing:1983qx}. We observe that the basis can be written in terms of ``towers'' attached to two ``seeds''. The first tower is simply the parametrisation of a scalar operator multiplied by the average four-velocity $P^\mu/M$ playing the role of seed. It can then naturally be interpreted as the convective part of the vector current. The second tower is associated with the seed proportional to $iS^{\mu\nu}\Delta_\nu$, where $S^{\mu\nu}$ is the generator of Lorentz transformations in either the four-vector [$(S^{\mu\nu})^{\alpha'\alpha}=i(g^{\alpha'\mu}g^{\nu\alpha}-g^{\alpha'\nu}g^{\mu\alpha})$] or the Dirac [$S^{\mu\nu}=\frac{1}{2}\sigma^{\mu\nu}$] representation. It can accordingly be interpreted as the spin or magnetisation part of the vector current~\cite{Schwartz:1955aa,Donnelly:1984rg}. The set of $2j+1$ vector FFs can therefore be decomposed into a set of $\lfloor j\rfloor+1$ convective FFs and a set of $\lceil j\rceil$ spin or magnetization FFs, where $\lceil j\rceil$ is the ceiling of the spin, i.e. the smallest integer greater or equal to $j$. Note that both convective and spin (or magnetisation) parts are separately conserved as one can easily check by contraction with $\Delta^\mu$. There is no way to construct a non-conserved Lorentz structure satisfying all the spacetime symmetry constraints.

For spin-$0$, one has:
\begin{equation}
\langle p',\lambda'|\hat J^\mu(0)|p,\lambda\rangle={\blue 2}P^\mu\,F^V_{1,0}(t).
\end{equation}

For spin-$\frac{1}{2}$:
\begin{equation}
\langle p',\lambda'|\hat J^\mu(0)|p,\lambda\rangle=\overline u(p',\lambda')\left[{\blue 2}P^\mu\,F^V_{1,0}(t)+{\blue i}\sigma^{\mu\nu}\Delta_\nu\,\,F^V_{2,0}(t)\right]u(p,\lambda).
\end{equation}

For spin-$1$:
\begin{equation}
\langle p',\lambda'|\hat J^\mu(0)|p,\lambda\rangle=-\varepsilon^*_{\alpha'}(p',\lambda')\left[{\blue 2}P^ \mu\left(g^{\alpha'\alpha}\,F^V_{1,0}(t)-\frac{\Delta^{\alpha'}\Delta^\alpha}{{\blue 2M^2}}\,F^V_{1,1}(t)\right)-\left(g^{\mu\alpha'}\Delta^{\alpha}-g^{\mu\alpha}\Delta^{\alpha'}\right)F^V_{2,0}(t)\right]\varepsilon_{\alpha}(p,\lambda).
\end{equation}

For spin-$\frac{3}{2}$:
\begin{equation}
\begin{aligned}
\langle p',\lambda'|\hat J^\mu(0)|p,\lambda\rangle&=-\overline u_{\alpha'}(p',\lambda')\left[{\blue 2}P^ \mu\left(g^{\alpha'\alpha}\,F^V_{1,0}(t)-\frac{\Delta^{\alpha'}\Delta^\alpha}{{\blue 2M^2}}\,F^V_{1,1}(t)\right)\right.\\
&\quad\left.+{\blue i}\sigma^{\mu\nu}\Delta_\nu\left(g^{\alpha'\alpha}\,F^V_{2,0}(t)-\frac{\Delta^{\alpha'}\Delta^\alpha}{{\blue 2M^2}}\,F^V_{2,1}(t)\right)\right]u_{\alpha}(p,\lambda).
\end{aligned}
\end{equation}

For spin-$2$:
\begin{equation}
\begin{aligned}
\langle p',\lambda'|\hat J^\mu(0)|p,\lambda\rangle&=\varepsilon^*_{\alpha'_1\alpha'_2}(p',\lambda')\left[{\blue 2}P^\mu\left(g^{\alpha'_1\alpha_1}g^{\alpha'_2\alpha_2}\,F^V_{1,0}(t)-\frac{\Delta^{\alpha'_1}\Delta^{\alpha_1}}{\blue 2M^2}\,g^{\alpha'_2\alpha_2}\,F^V_{1,1}(t)+\frac{\Delta^{\alpha'_1}\Delta^{\alpha_1}}{\blue 2M^2}\,\frac{\Delta^{\alpha'_2}\Delta^{\alpha_2}}{\blue 2M^2}\,F^V_{1,2}(t)\right)\right.\\
&\quad\left.-\left(g^{\mu\alpha'_2}\Delta^{\alpha_2}-g^{\mu\alpha_2}\Delta^{\alpha'_2}\right)\left(g^{\alpha'_1\alpha_1}\,F^V_{2,0}(t)-\frac{\Delta^{\alpha'_1}\Delta^{\alpha_1}}{\blue 2M^2}\,F^V_{2,1}(t)\right)\right]\varepsilon_{\alpha_1\alpha_2}(p,\lambda).
\end{aligned}
\end{equation}

\subsection{Tensor operator}

The last case we will treat explicitly is the tensor operator $\hat T^{\mu\nu}(x)$. A typical example is the energy-momentum tensor (EMT) $\hat T^{\mu\nu}_q(x)=\hat{\overline\psi}(x)\gamma^\mu iD^\nu\hat\psi(x)$. Various parametrisations have been proposed in the literature for spin-$0$~\cite{Pagels:1966zza,Donoghue:1991qv}, $\frac{1}{2}$~\cite{Kobzarev:1962wt,Ji:1996ek,Bakker:2004ib,Leader:2013jra}, and $1$~\cite{Holstein:2006ud,Abidin:2008ku,Taneja:2011sy,Cosyn:2019aio,Polyakov:2019lbq}. For the higher-spin cases, a few past works have investigated the rank-2 tensor with different results for the parametrisations, see e.g.~\cite{Cheshkov1963,Scadron:1968zz,Williams:1970ms,Chung:2019yfs}.  When $j=n$ is integer, we find that its elastic matrix elements can be written in terms of the following basis
\begin{equation}
\begin{aligned}
T^{\mu\nu,\alpha'_1\cdots\alpha'_n\alpha_1\cdots\alpha_n}(P,\Delta)&={\blue 2}P^\mu P^\nu\sum_{(k,n)}F^T_{1,k}(t)\\
&+{\blue 2}\left(\Delta^\mu\Delta^\nu-g^{\mu\nu}\Delta^2\right)\sum_{(k,n)}F^T_{2,k}(t)\\
&+ {\blue 2}M^2g^{\mu\nu}\sum_{(k,n)}F^T_{3,k}(t)\\
&-P^{\{\mu}g^{\nu\}[\alpha'_n}\Delta^{\alpha_n]}\sum_{(k,n-1)}F^T_{4,k}(t)\\
&-\left(\Delta^{\{\mu}g^{\nu\}\{\alpha'_n}\Delta^{\alpha_n\}}-g^{\mu\nu}\Delta^{\alpha'_n}\Delta^{\alpha_n}-g^{\alpha'_n\{\mu}g^{\nu\}\alpha_n}\Delta^2\right)\sum_{(k,n-1)}F^T_{5,k}(t)\\
&+M^2g^{\alpha'_n\{\mu}g^{\nu\}\alpha_n}\sum_{(k,n-1)}F^T_{6,k}(t)\\
&+\Delta^{[\alpha'_n}g^{\alpha_n]\{\mu}g^{\nu\}[\alpha'_{n-1}}\Delta^{\alpha_{n-1}]}\sum_{(k,n-2)}F^T_{7,k}(t)\\
&-P^{[\mu}g^{\nu][\alpha'_n}\Delta^{\alpha_n]}\sum_{(k,n-1)}F^T_{8,k}(t)\\
&-\Delta^{[\mu}g^{\nu]\{\alpha'_n}\Delta^{\alpha_n\}}\sum_{(k,n-1)}F^T_{9,k}(t),
\end{aligned}
\end{equation}
where $a^{\{\mu}b^{\nu\}}=a^\mu b^\nu+a^\nu b^\mu$ and $a^{[\mu}b^{\nu]}=a^\mu b^\nu-a^\nu b^\mu$. When $j=n+\frac{1}{2}$ is half-integer, we find a similar basis
\begin{equation}
\begin{aligned}
T^{\mu\nu,\alpha'_1\cdots\alpha'_n\alpha_1\cdots\alpha_n}(P,\Delta)&={\blue 2}P^\mu P^\nu\sum_{(k,n)}F^T_{1,k}(t)\\
&+{\blue 2}\left(\Delta^\mu\Delta^\nu-g^{\mu\nu}\Delta^2\right)\sum_{(k,n)}F^T_{2,k}(t)\\
&+{\blue 2}M^2g^{\mu\nu}\sum_{(k,n)}F^T_{3,k}(t)\\
&+P^{\{\mu}\tfrac{i}{2}\sigma^{\nu\}\rho}\Delta_\rho\sum_{(k,n)}F^T_{4,k}(t)\\
&-\left(\Delta^{\{\mu}g^{\nu\}\{\alpha'_n}\Delta^{\alpha_n\}}-g^{\mu\nu}\Delta^{\alpha'_n}\Delta^{\alpha_n}-g^{\alpha'_n\{\mu}g^{\nu\}\alpha_n}\Delta^2\right)\sum_{(k,n-1)}F^T_{5,k}(t)\\
&+M^2g^{\alpha'_n\{\mu}g^{\nu\}\alpha_n}\sum_{(k,n-1)}F^T_{6,k}(t)\\
&+\Delta^{[\alpha'_n}g^{\alpha_n]\{\mu}g^{\nu\}[\alpha'_{n-1}}\Delta^{\alpha_{n-1}]}\sum_{(k,n-2)}F^T_{7,k}(t)\\
&+P^{[\mu}\tfrac{i}{2}\sigma^{\nu]\rho}\Delta_\rho\sum_{(k,n)}F^T_{8,k}(t)\\
&-\Delta^{[\mu}g^{\nu]\{\alpha'_n}\Delta^{\alpha_n\}}\sum_{(k,n-1)}F^T_{9,k}(t).
\label{tensor_tower}
\end{aligned}
\end{equation}
The total number of tensor FFs is:
\begin{equation}
   4j+5\lfloor j\rfloor+3-\theta(j\geq 1), 
\end{equation}
where we defined $\theta(j\geq 1)=1$ when $j\geq 1$, and zero otherwise. In particular, the symmetric conserved part (associated to the FFs $F_{i,k}^T$ with $i=1,2,4,5,7$) is parametrised in terms of $2(j+1)+3\lfloor j\rfloor-\theta(j\geq 1)$ FFs. The remaining FFs in the parametrisation are divided as follows: $2\lfloor j\rfloor+1$ FFs come  from the symmetric non-conserved part ($i=3,6$),  $\lceil j\rceil$ of them are related to the antisymmetric conserved part ($i=8$), and the last $\lfloor j\rfloor$ FFs come from the antisymmetric non-conserved part ($i=9$).
 This agrees with former results for spin $0$, $\tfrac{1}{2}$, and $1$. We arranged the bases so to maximise the number of conserved terms. This is especially important in view of the application to the EMT, where non-conserved terms play a key role in identifying separate quark and gluon contributions~\cite{Lorce:2017xzd,Lorce:2018egm}. 

For spin-$0$, $\frac{1}{2}$, and $1$ we recover the known parametrisations. For spin-$0$, one has:
\begin{equation}
\langle p',\lambda'|\hat T^{\mu\nu}(0)|p,\lambda\rangle=
{\blue 2}P^\mu P^\nu\,F^T_{1,0}(t)
+{\blue 2}\left(\Delta^\mu\Delta^\nu-g^{\mu\nu}\Delta^2\right)F^T_{2,0}(t)
+{\blue 2}M^2g^{\mu\nu} F^T_{3,0}(t).
\end{equation}

For spin-$\frac{1}{2}$:
\begin{equation}
\begin{aligned}
\langle p',\lambda'|\hat T^{\mu\nu}(0)|p,\lambda\rangle &= \overline u(p',\lambda')\Big[
 {\blue 2}P^\mu P^\nu\,F^T_{1,0}(t)
+{\blue 2}\left(\Delta^\mu\Delta^\nu-g^{\mu\nu}\Delta^2\right)F^T_{2,0}(t)\\ &
\quad
+{\blue 2}M^2g^{\mu\nu} F^T_{3,0}(t)+P^{\{\mu}\tfrac{i}{2}\sigma^{\nu\}\rho}\Delta_\rho F^T_{4,0}(t)+P^{[\mu}\tfrac{i}{2}\sigma^{\nu]\rho}\Delta_\rho F^T_{8,0}(t)
\Big]u(p,\lambda).
\end{aligned}
\end{equation}

For spin-$1$:
\begin{equation}
\begin{aligned}
\langle p',\lambda'|\hat T^{\mu\nu}(0)|p,\lambda\rangle&=-\varepsilon^*_{\alpha'}(p',\lambda')\bigg[
{\blue 2}P^ \mu P^\nu \left(g^{\alpha'\alpha}\,F^T_{1,0}(t)-\frac{\Delta^{\alpha'}\Delta^\alpha}{\blue 2M^2}\,F^T_{1,1}(t)\right)\\
&\quad +{\blue 2}\left(\Delta^\mu\Delta^\nu-g^{\mu\nu}\Delta^2\right)\left(g^{\alpha'\alpha}\,F^T_{2,0}(t)-\frac{\Delta^{\alpha'}\Delta^\alpha}{\blue 2M^2}\,F^T_{2,1}(t)\right)+{\blue 2}M^2g^{\mu\nu}\left(g^{\alpha'\alpha}\,F^T_{3,0}(t)-\frac{\Delta^{\alpha'}\Delta^\alpha}{\blue 2M^2}\,F^T_{3,1}(t) \right)\\
&\quad -P^{\{\mu}g^{\nu\}[\alpha'}\Delta^{\alpha]}F^T_{4,0}(t)-\left(\Delta^{\{\mu}g^{\nu\}\{\alpha'}\Delta^{\alpha\}}-g^{\mu\nu}\Delta^{\alpha'}\Delta^{\alpha}-g^{\alpha'\{\mu}g^{\nu\}\alpha}\Delta^2\right)F^T_{5,0}(t)\\
&\quad +M^2g^{\alpha'\{\mu}g^{\nu\}\alpha} F^T_{6,0}(t)-P^{[\mu}g^{\nu][\alpha'}\Delta^{\alpha]}F^T_{8,0}(t) -\Delta^{[\mu}g^{\nu]\{\alpha'}\Delta^{\alpha\}}F^T_{9,0}(t)\bigg]\varepsilon_{\alpha}(p,\lambda).
\end{aligned}
\end{equation}

For spin-$\frac{3}{2}$:
\begin{equation}
\begin{aligned}
\langle p',\lambda'|\hat T^{\mu\nu}(0)|p,\lambda\rangle&=-\overline u_{\alpha'}(p',\lambda')\bigg[
{\blue 2}P^ \mu P^\nu \left(g^{\alpha'\alpha}\,F^T_{1,0}(t)-\frac{\Delta^{\alpha'}\Delta^\alpha}{\blue 2M^2}\,F^T_{1,1}(t)\right)\\
&\quad +{\blue 2}\left(\Delta^\mu\Delta^\nu-g^{\mu\nu}\Delta^2\right)\left(g^{\alpha'\alpha}\,F^T_{2,0}(t)-\frac{\Delta^{\alpha'}\Delta^\alpha}{\blue 2M^2}\,F^T_{2,1}(t)\right)+{\blue 2}M^2g^{\mu\nu}\left(g^{\alpha'\alpha}\,F^T_{3,0}(t)-\frac{\Delta^{\alpha'}\Delta^\alpha}{\blue 2M^2}\,F^T_{3,1}(t) \right)\\
&\quad +P^{\{\mu}\tfrac{i}{2}\sigma^{\nu\}\rho}\Delta_\rho \left(g^{\alpha'\alpha}\,F^T_{4,0}(t)-\frac{\Delta^{\alpha'}\Delta^\alpha}{\blue 2M^2}\,F^T_{4,1}(t)\right)\\
&\quad -\left(\Delta^{\{\mu}g^{\nu\}\{\alpha'}\Delta^{\alpha\}}-g^{\mu\nu}\Delta^{\alpha'}\Delta^{\alpha}-g^{\alpha'\{\mu}g^{\nu\}\alpha}\Delta^2\right)F^T_{5,0}(t) +M^2g^{\alpha'\{\mu}g^{\nu\}\alpha}F^T_{6,0}(t)\\
&\quad+P^{[\mu}\tfrac{i}{2}\sigma^{\nu]\rho}\Delta_\rho \left(g^{\alpha'\alpha}\,F^T_{8,0}(t)-\frac{\Delta^{\alpha'}\Delta^\alpha}{\blue 2M^2}\,F^T_{8,1}(t)\right) -\Delta^{[\mu}g^{\nu]\{\alpha'}\Delta^{\alpha\}}F^T_{9,0}(t)
\bigg]u_{\alpha}(p,\lambda).
\end{aligned}
\end{equation}

For spin-$2$:
\begin{equation}
\begin{aligned}
\langle p',\lambda'|\hat T^{\mu\nu}(0)|p,\lambda\rangle
&=\varepsilon^*_{\alpha'_1\alpha'_2}(p',\lambda')\bigg[
{\blue 2}P^\mu P^\nu\left(g^{\alpha'_1\alpha_1}g^{\alpha'_2\alpha_2}\,F^T_{1,0}(t)-\frac{\Delta^{\alpha'_1}\Delta^{\alpha_1}}{\blue 2M^2}\,g^{\alpha'_2\alpha_2}\,F^T_{1,1}(t)+\frac{\Delta^{\alpha'_1}\Delta^{\alpha_1}}{\blue 2M^2}\,\frac{\Delta^{\alpha'_2}\Delta^{\alpha_2}}{\blue 2M^2}\,F^T_{1,2}(t)\right)\\
&\quad+{\blue 2}\left(\Delta^\mu\Delta^\nu-g^{\mu\nu}\Delta^2\right)\left(g^{\alpha'_1\alpha_1}g^{\alpha'_2\alpha_2}\,F^T_{2,0}(t)-\frac{\Delta^{\alpha'_1}\Delta^{\alpha_1}}{\blue 2M^2}\,g^{\alpha'_2\alpha_2}\,F^T_{2,1}(t)+\frac{\Delta^{\alpha'_1}\Delta^{\alpha_1}}{\blue 2M^2}\,\frac{\Delta^{\alpha'_2}\Delta^{\alpha_2}}{\blue 2M^2}\,F^T_{2,2}(t)\right)\\
&\quad+{\blue 2}M^2g^{\mu\nu}\left(g^{\alpha'_1\alpha_1}g^{\alpha'_2\alpha_2}\,F^T_{3,0}(t)-\frac{\Delta^{\alpha'_1}\Delta^{\alpha_1}}{\blue 2M^2}\,g^{\alpha'_2\alpha_2}\,F^T_{3,1}(t)+\frac{\Delta^{\alpha'_1}\Delta^{\alpha_1}}{\blue 2M^2}\,\frac{\Delta^{\alpha'_2}\Delta^{\alpha_2}}{\blue 2M^2}\,F^T_{3,2}(t)\right)\\
&\quad-P^{\{\mu}g^{\nu\}[\alpha'_2}\Delta^{\alpha_2]}\left(g^{\alpha'_1\alpha_1}F^T_{4,0}(t)-\frac{\Delta^{\alpha'_1}\Delta^{\alpha_1}}{\blue 2M^2}F^T_{4,1}(t)\right)\\
&\quad-\left(\Delta^{\{\mu}g^{\nu\}\{\alpha'_2}\Delta^{\alpha_2\}}-g^{\mu\nu}\Delta^{\alpha'_2}\Delta^{\alpha_2}-g^{\alpha'_2\{\mu}g^{\nu\}\alpha_2}\Delta^2\right)\left(g^{\alpha'_1\alpha_1}F^T_{5,0}(t)-\frac{\Delta^{\alpha'_1}\Delta^{\alpha_1}}{\blue 2M^2}F^T_{5,1}(t)\right)\\
&\quad+M^2g^{\alpha'_2\{\mu}g^{\nu\}\alpha_2} \left(g^{\alpha'_1\alpha_1}F^T_{6,0}(t)-\frac{\Delta^{\alpha'_1}\Delta^{\alpha_1}}{\blue 2M^2}F^T_{6,1}(t)\right) +\Delta^{[\alpha'_2}g^{\alpha_2]\{\mu}g^{\nu\}[\alpha'_{1}}\Delta^{\alpha_{1}]}F^T_{7,0}(t)\\
&\quad-P^{[\mu}g^{\nu][\alpha'_2}\Delta^{\alpha_2]}\left(g^{\alpha'_1\alpha_1}F^T_{8,0}(t)-\frac{\Delta^{\alpha'_1}\Delta^{\alpha_1}}{\blue 2M^2}F^T_{8,1}(t)\right)\\
&\quad-\Delta^{[\mu}g^{\nu]\{\alpha'_2}\Delta^{\alpha_2\}}\left(g^{\alpha'_1\alpha_1}F^T_{9,0}(t)-\frac{\Delta^{\alpha'_1}\Delta^{\alpha_1}}{\blue 2M^2}F^T_{9,1}(t)\right)
\bigg]\varepsilon_{\alpha_1\alpha_2}(p,\lambda).
\end{aligned}
\end{equation}

\section{Multipole expansion technique}\label{sec:multipole}

The tensor product approach used in the previous section has the advantage of giving the explicit form for all the possible structures in the given representation. We explicitly derived the parametrisations up to spin 2, which is necessary in particular for the tensor operator case. A further increase in the target spin has the effect of introducing additional factors of $g^{\alpha_i^\prime\alpha_i}$ and $\Delta^{\alpha_i^\prime}\Delta^{\alpha_i}$, feeding the towers with new elements, as is clear from the notation~\eqref{strange_sum}. The main disadvantage however is that it is based on the direct inspection of the structures and on the explicit use of several on-shell identities which are highly non-trivial. This is the reason why several former parametrisations proposed in the literature have been found either incomplete or overcomplete. 

In the following, we develop another technique based on covariant multipoles, complementary to the tensor product approach and confirming the number of FFs. As already stressed earlier, the choice of basis for a parametrisation is arbitrary. Some bases appear however to be more useful because of their mathematical simplicity or their physical meaning. The multipole basis is especially interesting since it is closely related to the symmetries of the problem. In particular, it clarifies how parametrisations associated with different target spins are related to each other.

\subsection{Standard $\mathfrak{su}(2)$ multipoles}

In non-relativistic descriptions, it is often convenient to expand physical quantities in terms of multipoles associated with the three-dimensional rotation group. In relativistic descriptions, the rotation group appears as the little group associated with massive representations, i.e. the subgroup of the Lorentz group which leaves the (timelike) four-momentum $p^\mu$ invariant. Accordingly, the standard $\mathfrak{su}(2)$ multipole expansion remains useful as long as no four-momentum is transferred to the system. This explains for example why one can use essentially the same multipole expansion for the spin density matrix in both non-relativistic and relativistic descriptions~\cite{Leader:2001gr}.

In a given spin representation, operators can conveniently be expanded into products of the rotation generators $J^i$
\begin{equation}
O=cI+c^iJ^i+c^{ij}J^iJ^j+\cdots,
\end{equation}
where $I$ is the identity and $c^{ij\cdots}$ are $\mathds C$-valued coefficients. Because of the $\mathfrak{su}(2)$ Lie algebra $[J^i,J^j]=i\epsilon^{ijk}J^k$ and the $\mathfrak{su}(2)$ Casimir $\boldsymbol J^2=j(j+1)I$, the coefficients of the multipole expansion can be taken completely symmetric and traceless, and the $\mathfrak{su}(2)$ multipoles of order $k$ are defined as
\begin{equation}
M_k^{i_1\cdots i_k}\equiv\mathds S J^{i_1}\cdots J^{i_k},    
\end{equation}
where $\mathds S$ indicates that the product is symmetrised and traceless. For example, the first three multipoles read:
\begin{equation}
\begin{aligned}
\text{Monopole}&&M_0&=I,\\
\text{Dipole}&&M_1^i&=J^i,\\
\text{Quadrupole}&&M_2^{ij}&=\frac{1}{2}\{J^i,J^j\}-\frac{1}{3}\,\delta^{ij}\boldsymbol J^2.
\end{aligned}
\end{equation}
One can then write the multipole expansion as
\begin{equation}
O=\sum_k c^{i_1\cdots i_k} M^{i_1\cdots i_k}_k.
\end{equation}
Since the spin representation has finite dimension, the Cayley-Hamilton theorem ensures that the expansion must stop at some finite order~\cite{Lang1970}. More precisely, a spin-$j$ representation will admit only the first $2j+1$ multipoles. Multipoles of higher order simply vanish.

\subsection{Covariant $\mathfrak{sl}(2,\mathds C)$ multipoles}

The rotation group being a subgroup of the Lorentz group motivates the extension of the multipole expansion technique to the whole set of $\mathfrak{sl}(2,\mathds C)$ generators $S^{\mu\nu}$. That such an expansion exists has been suggested by a complete parametrisation of the EMT for spin-1 hadrons, see Appendices B and C of~\cite{Cosyn:2019aio}. Recently, elements of a covariant multipole expansion have been exposed in~\cite{Bautista:2019tdr}. Our aim here is to develop this technique further before applying it to our problem.

Similarly to the $\mathfrak{su}(2)$ case, one can conveniently expand operators in a given representation into products of the Lorentz generators
\begin{equation}\label{multexp0}
O=cI+c_{\mu\nu}S^{\mu\nu}+c_{\mu\nu,\rho\sigma}S^{\mu\nu}S^{\rho\sigma}+\cdots
\end{equation}
In the following, we will refer to pairs of antisymmetric indices as bi-indices. Because of the $\mathfrak{sl}(2,\mathds C)$ Lie algebra $[S^{\mu\nu},S^{\rho\sigma}]=i(g^{\mu\rho}S^{\nu\sigma}-g^{\nu\rho}S^{\mu\sigma}+g^{\mu\sigma}S^{\rho\nu}-g^{\nu\sigma}S^{\rho\mu})$, the coefficients of the multipole expansion can be taken completely symmetric under the exchange of bi-indices
\begin{equation}
c_{\cdots,\mu\nu,\cdots,\rho\sigma,\cdots}=c_{\cdots,\rho\sigma,\cdots,\mu\nu,\cdots}
\end{equation} 
Unlike the $\mathfrak{su}(2)$ case, there are two types of ``traces'' for bi-indices. The first one, $g_{\mu[\rho}g_{\sigma]\nu}$, has mixed symmetry and is even under parity, whereas the second one, $i\epsilon_{\mu\nu\rho\sigma}$, is totally antisymmetric and odd under parity. Accordingly one gets two quadratic $\mathfrak{sl}(2,\mathds C)$ Casimirs $C\equiv\frac{1}{2}S^{\mu\nu}S_{\mu\nu}=\boldsymbol J^2-\boldsymbol K^2=2[j_1(j_1+1)+j_2(j_2+1)]$ and $\tilde C\equiv\frac{i}{4}\epsilon_{\mu\nu\rho\sigma}S^{\mu\nu}S^{\rho\sigma}=i\{J^i,K^i\}=2[j_1(j_1+1)-j_2(j_2+1)]$. For example, for a Dirac particle one gets $C\propto\mathds 1$ and $\tilde C\propto\gamma_5$, both matrices commuting indeed with the Lorentz generators $S^{\mu\nu}=\frac{1}{2}\sigma^{\mu\nu}$. The four-vector and more generally all $(\frac{n}{2},\frac{n}{2})$ representations are characterised by the vanishing of the second Casimir $\tilde C=0$. The multipole expansion~\eqref{multexp0} can then be reorganised as follows (the coefficients $c$ and $\tilde c$ do not involve the Levi-Civita tensor)
\begin{equation}\label{multexp1}
    O=\sum_k (c_{\mu_1\nu_1,\cdots,\mu_k\nu_k}I+\tilde c_{\mu_1\nu_1,\cdots,\mu_k\nu_k}\tilde C)\,\mathcal M_k^{\mu_1\nu_1,\cdots,\mu_k\nu_k},
\end{equation}
where the $\mathfrak{sl}(2,\mathds C)$ multipoles of order $k$ are defined as 
\begin{equation}
\mathcal M_k^{\mu_1\nu_1,\cdots, \mu_k\nu_k}\equiv\mathds S S^{\mu_1\nu_1}\cdots S^{\mu_k\nu_k},    
\end{equation}
with $\mathds S$ indicating here that the product is symmetrised over bi-indices and all traces are removed. The latter involve two, three or four bi-indices
\begin{equation}
    \begin{aligned}
 \mathcal M^{\cdots,\mu\nu,\cdots,\rho\sigma,\cdots}g_{\mu\nu\rho\sigma}&=0,\\
 \mathcal M^{\cdots,\mu\mu',\cdots,\nu\nu',\cdots,\rho\sigma,\cdots}g_{\mu\nu\rho\sigma}&=0,\\ 
 \mathcal M^{\cdots,\mu\mu',\cdots,\nu\nu',\cdots,\rho\rho',\cdots,\sigma\sigma',\cdots}g_{\mu\nu\rho\sigma}&=0,
    \end{aligned}
\end{equation}
where $g_{\mu\nu\rho\sigma}=g_{\mu[\rho}g_{\sigma]\nu}$ or $i\epsilon_{\mu\nu\rho\sigma}$. For example, the first three covariant multipoles read:
\begin{equation}
\begin{aligned}
\text{Monopole}&&\mathcal M_0&=I,\\
\text{Dipole}&&\mathcal M_1^{\mu\nu}&=S^{\mu\nu},\\
\text{Quadrupole}&&\mathcal M_2^{\mu\nu,\rho\sigma}&=\frac{1}{2}\{S^{\mu\nu},S^{\rho\sigma}\}-\frac{1}{12}\,g^{\mu[\rho}g^{\sigma]\nu}S^{\lambda\tau}S_{\lambda\tau}+\frac{1}{4!}\epsilon^{\mu\nu\rho\sigma}\epsilon_{\lambda\tau\lambda'\tau'}S^{\lambda\tau}S^{\lambda'\tau'}.
\end{aligned}
\end{equation}
Once again Cayley-Hamilton theorem ensures that in a spin-$j$ representation the expansion stops at order $k=2j$.

\section{Parametrisation in terms of covariant multipoles}\label{sec:covariantparam}

We construct now an alternative parametrisation for the matrix elements of the scalar, vector and tensor operators in terms of covariant multipoles. Like in Section~\ref{sec:param1} we will use the $(\frac{n}{2},\frac{n}{2})$ representation for integer spin targets and the $(\frac{n+1}{2},\frac{n}{2})\oplus(\frac{n}{2},\frac{n+1}{2})$ representation for half-integer spin targets. Using various on-shell relations derived in Appendix~\ref{AppB}, we observe that all covariant multipoles contracted with $P^\mu$ can be discarded from the list of independent tensor structures. Discrete symmetries expressed by the constraints in Eqs.~\eqref{dissymm1} and~\eqref{dissymm2} allow us to further reduce the number of independent tensor structures. Since we are restricting ourselves to operators with positive intrinsic parity, we can set $\tilde c_{\mu_1\nu_1,\cdots,\mu_k\nu_k}=0$ in the multipole expansion~\eqref{multexp1}. Time-reversal symmetry implies that odd covariant multipoles should be multiplied by $i$ for the FFs to be real-valued, and hermiticity imposes that the coefficients in front of even (odd) multipoles involve an even (odd) number of factors of $\Delta^\mu$. 

\subsection{Scalar operator}

In the case of a scalar operator, we need to fully contract the bi-indices of the multipoles. Owing to the above constraints, only even multipoles can be used
\begin{equation}\label{canonicalform}
\begin{aligned}
&\mathcal M_0,\\
&\mathcal M^{\rho_1\Delta,\rho_2\Delta}_2g_{\rho_1\rho_2},\\
&\mathcal M^{\rho_1\Delta,\rho_2\Delta,\rho_3\Delta,\rho_4\Delta}_4g_{\rho_1\rho_2}g_{\rho_3\rho_4},\\
&\,\,\vdots
 \end{aligned}
\end{equation}
where an index $\Delta$ means contraction with $\Delta^\sigma$, e.g. $\mathcal M ^{\rho\Delta,\cdots}=\mathcal M^{\rho\sigma,\cdots}\Delta_\sigma$. Note that since covariant multipoles are symmetric under the exchange of bi-indices, the independent contractions can always be put in the canonical form~\eqref{canonicalform}. For convenience we shall use the notation
\begin{equation}\label{bullet_notation}
\mathcal M^{\cdots,\bullet\Delta,\bullet\Delta}_{2k}=\mathcal M^{\cdots,\rho_1\Delta,\rho_2\Delta}_{2k}g_{\rho_1\rho_2}.
\end{equation}
Since a multipole $\mathcal M_{2k}$ has $2k$ bi-indices, the fully contracted even multipoles of the scalar parametrisation contain $k$ pairwise contractions. The fact that there is only one type of contraction associated with each multipole is reflected in the existence of only one seed in the parametrisation~\eqref{scalarop}.

The multipole parametrisation for the scalar operator then reads
\begin{equation}
N(P,\Delta)={\blue 2M}\sum_{k}\frac{1}{{\blue 2^k}M^{2k}}\,\mathcal M^{\bullet\Delta,\cdots,\bullet\Delta}_{2k}\,\mathcal F^S_k(t),
\end{equation}
where $k$ runs over integer numbers and the series truncates for $k>j$, because the (even) multipoles $\mathcal M_{2k}$ vanishes. Increasing the multipole order with the same contraction pattern generates the tower associated by the strange sum multiplying the seed in~\eqref{scalarop}. Since only even multipoles are allowed, this explains why the number of scalar FFs is $\lfloor j\rfloor+1$.

\subsection{Vector operator}

A vector operator carries one open Lorentz index. In constructing the multipole decomposition, we bare in mind that the open index is carried either by the coefficient or by the multipole. In the first case we are left again with only even multipoles, whereas in the second case only odd multipoles can contribute, i.e.
\begin{equation}
\begin{aligned}
&{\blue 2}P^\mu\, \mathcal M_0,\\
&i\mathcal M^{\mu\Delta}_1,\\
&{\blue 2}P^\mu\, \mathcal M^{\bullet\Delta,\bullet\Delta}_2,\\
&i\mathcal M^{\mu\Delta,\bullet\Delta,\bullet\Delta}_3,\\
&{\blue 2}P^\mu\, \mathcal M^{\bullet\Delta,\bullet\Delta,\bullet\Delta,\bullet\Delta}_4,\\
&\,\,\vdots
 \end{aligned}
\end{equation}
In other words, there is an additional possible contraction which concerns only the odd multipoles. 

The multipole parametrisation for the vector current then reads
\begin{equation}
\begin{aligned}
J^\mu(P,\Delta)=\sum_{k}\frac{1}{{\blue 2^k}M^{2k}}&\left[{\blue 2}P^\mu\,\mathcal M^{\bullet\Delta,\cdots,\bullet\Delta}_{2k}\,\mathcal F^V_{1,k}(t)\right.\\
&\,\left.+i\mathcal M^{\mu\Delta,\bullet\Delta,\cdots,\bullet\Delta}_{2k+1}\,\mathcal F^V_{2,k}(t)\right].
\end{aligned}
\end{equation}
Here it is understood that $k$ runs over integers and that the series truncates when the $n$-th multipole $\mathcal M_n$ vanishes, i.e.\ for $n>2j$. The total number of terms is $2j+1$, where $\lfloor j\rfloor +1$ FFs come from the even multipole expansion and the remaining $\lceil j\rceil$ come from the odd multipole expansion. Each term is manifestly conserved. Note that thanks to the covariant multipole approach, we are able to write a single parametrisation valid for both integer and half-integer spins, as already suggested by the similitude between Eqs.~\eqref{Jmuint} and~\eqref{Jmuhalfint}.

\subsection{Tensor operator}

For a tensor operator, the two open Lorentz indices can be carried entirely by the coefficient, or the multipole, or both. In the first two cases only even multipoles appear, whereas in the last case both even and odd multipoles contribute. 
In particular, new types of contractions of the even multipoles appear in addition to those involved for the vector operator. They can be put in the canonical form
\begin{equation}
 \mathcal M_{2k}^{\mu\bullet,\nu\bullet,\bullet\Delta,\cdots,\bullet\Delta},\quad \Delta^{\{\mu}\mathcal M_{2k}^{\nu\}\bullet,\bullet\Delta,\cdots,\bullet\Delta},\quad \Delta^{[\mu}\mathcal M_{2k}^{\nu]\bullet,\bullet\Delta,\cdots,\bullet\Delta},\quad   \mathcal M_{2k}^{\mu\Delta,\nu\Delta,\bullet\Delta,\cdots,\bullet\Delta}
\end{equation}
owing to the symmetry under the exchange of bi-indices and the tracelessness of the covariant mulipoles. These operators have $k$ pairwise contractions, except for $\mathcal M_{2k}^{\mu\Delta,\nu\Delta,\bullet\Delta,\cdots,\bullet\Delta}$ which has $k-1$ contractions and which is responsible for generating the tower associated with the FFs $F^T_{k,7}$ of Eq.~\eqref{tensor_tower}. It appears only for $0<k<\lfloor j\rfloor$, while for $k=\lfloor j\rfloor$ it is not independent of the other types of contractions involving the multipole $\mathcal M_{2\lfloor j\rfloor}$. For example, for spin $j<2$ we find that
\begin{equation}
2\, \mathcal M_{2}^{\mu\Delta,\nu\Delta,\bullet\Delta,\cdots,\bullet\Delta}-\Delta^{\{\mu}\mathcal M_{2}^{\nu\}\bullet,\bullet\Delta,\cdots,\bullet\Delta}-\Delta^2\,\mathcal M_{2}^{\mu\bullet,\nu\bullet,\bullet\Delta,\cdots,\bullet\Delta}-g^{\mu\nu}\,\mathcal M_{2}^{\bullet\Delta,\cdots,\bullet\Delta}=0.
\label{identityQuadr}
\end{equation}

The multipole parametrisation for the tensor current maximising the number of conserved structures then reads
\begin{equation}
\begin{aligned}
T^{\mu\nu}(P,\Delta)=\sum_k\frac{1}{{\blue 2^k}M^{2k}}&\left[{\blue 2}P^\mu P^\nu\,\mathcal M^{\bullet\Delta,\cdots,\bullet\Delta}_{2k}\,\mathcal F^T_{1,k}(t)\right.\\
&\,+{\blue 2}\left(\Delta^\mu\Delta^\nu-g^{\mu\nu}\Delta^2\right)\mathcal M^{\bullet\Delta,\cdots,\bullet\Delta}_{2k}\,\mathcal F^T_{2,k}(t)\\
&\,+{\blue 2}M^2 g^{\mu\nu}\,\mathcal M^{\bullet\Delta,\cdots,\bullet\Delta}_{2k}\,\mathcal F^T_{3,k}(t)\\
&\,+P^{\{\mu}i\mathcal M^{\nu\}\Delta,\bullet\Delta,\cdots,\bullet\Delta}_{2k+1}\,\mathcal F^T_{4,k}(t)\\
&\,+\left(\Delta^{\{\mu}\mathcal M_{2k}^{\nu\}\bullet,\bullet\Delta,\cdots,\bullet\Delta}+\Delta^2\,\mathcal M_{2k}^{\mu\bullet,\nu\bullet,\bullet\Delta,\cdots,\bullet\Delta}+g^{\mu\nu}\,\mathcal M_{2k}^{\bullet\Delta,\cdots,\bullet\Delta}\right)\mathcal F^T_{5,k}(t)\\
&\,+M^2\, \mathcal M_{2k}^{\mu\bullet,\nu\bullet,\bullet\Delta,\cdots,\bullet\Delta}\,\mathcal F^T_{6,k}(t)\\
&\,+\theta(\lfloor j\rfloor>k)\,\mathcal M_{2k}^{\mu\Delta,\nu\Delta,\bullet\Delta,\cdots,\bullet\Delta}\,\mathcal F^T_{7,k}(t)\\
&\,+P^{[\mu}i\mathcal M^{\nu]\Delta,\bullet\Delta,\cdots,\bullet\Delta}_{2k+1}\,\mathcal F^T_{8,k}(t)\\
&\,\left.+\Delta^{[\mu}\mathcal M_{2k}^{\nu]\bullet,\bullet\Delta,\cdots,\bullet\Delta}\,\mathcal F^T_{9,k}(t)\right].
\label{tensor_multipole}
\end{aligned}
\end{equation}
Here it is again understood that $k$ runs over integers and that the series truncates when the $n$-th multipole $\mathcal M_n$ vanishes, i.e.\ for $n>2j$. Contrary to the scalar and vector cases, the relations between the ``curly'' tensor FFs $\mathcal F^T_{i,k}$ of this section and the ``straight'' $F^T_{j,k}$ of Section~\ref{sec:param1} mix in general different towers $i\neq j$. This can easily be seen by writing down explicitly the covariant multipoles in the $(\frac{n}{2},\frac{n}{2})$ and $(\frac{n+1}{2},\frac{n}{2})\oplus(\frac{n}{2},\frac{n+1}{2})$ representations. In the symmetric part, there are $3(\lfloor j\rfloor+1)$ FFs associated with fully contracted multipoles ($i=1,2,3$), $3\lfloor j\rfloor-\theta(j\geq 1)$ FFs associated with partially contracted even multipoles ($i=5,6,7$), and $\lceil j\rceil$ FFs associated with odd multipoles ($i=4$). In the antisymmetric part, there are $\lfloor j\rfloor$ FFs associated with partially contracted even multipoles ($i=9$) and $\lceil j\rceil$ FFs associated with odd multipoles ($i=8$). The total number of FFs is therefore $4j+5\lfloor j\rfloor+3-\theta(j\geq 1)$ in agreement with the counting of Section~\ref{sec:param1}.

In Refs.~\cite{Cotogno:2019xcl,Lorce:2019sbq} general constraints from Poincar\'e symmetry have been derived for targets with arbitrary spin. The key point was that the symmetric part of the conserved total EMT involves only two Lorentz structures to linear order in $\Delta$. Expanding our complete parametrisation~\eqref{tensor_multipole} to that order, we find
\begin{equation}
\begin{aligned}
T^{\mu\nu}(P,\Delta)={\blue 2}P^\mu P^\nu\,\mathcal F^T_{1,0}(0)+{\blue 2}M^2 g^{\mu\nu}\,\mathcal F^T_{3,0}(0)+P^{\{\mu} iS^{\nu\}\Delta}\,\mathcal F^T_{4,0}(0)+M^2\, \mathcal M_{2}^{\mu\bullet,\nu\bullet}\,\mathcal F^T_{6,1}(0)+P^{[\mu}iS^{\nu]\Delta}\,\mathcal F^T_{8,0}(0)+\mathcal O(\Delta^2).
\end{aligned}
\end{equation}
Keeping only the symmetric conserved part, we are left with ${\blue 2}P^\mu P^\nu\,\mathcal F^T_{1,0}(0)+P^{\{\mu} iS^{\nu\}\Delta}\,\mathcal F^T_{4,0}(0)$. Poincar\'e symmetry then imposes that for the total EMT $\mathcal F^T_{1,0}(0)=\mathcal F^T_{4,0}(0)=1$~\cite{Cotogno:2019xcl,Lorce:2019sbq}.

\subsection{Summary of the results for spin $j\le 2$}
In Table~\ref{Table_summary} we display the complete set of structures appearing in the parametrisations of scalar, vector, and tensor operators for spin values up to 2. The number of FFs for each spin value includes the structures at the relevant spin entry and all the ones above it. The results should be compared with Section~\ref{sec:param1}.

\begin{table}[h!]
\begin{tabular}{*5{>{\centering\arraybackslash}m{1.25in}} @{}m{0pt}@{}}
\hline
\bf{Spin}&\bf{Multipoles} & \bf{Scalar }$ N$ & \bf{Vector }$J^\mu$ & \bf{Tensor }$T^{\mu\nu}$& \\ [3.5ex] \hline
$j\geq 0$ &$\mathcal M_0$ & $\mathcal M_0$ & $P^\mu \mathcal M_0$ & \tabincell{c}{\vphantom{c}\\$P^\mu P^\nu \mathcal M_0$\\ $\Delta^\mu\Delta^\nu \mathcal M_0$\\ $g^{\mu\nu} \mathcal M_0 $\\ \vphantom{c}}  & \\ [3.5ex] \hline
$j\geq\frac{1}{2}$ & $\mathcal M_1^{\mu_1\nu_1}$ & $-$ & $i\mathcal M_1^{\mu\Delta}$ &\tabincell{c}{\vphantom{c}\\$P^{\{\mu} i\mathcal M_1^{\nu\}\Delta} $\\ $P^{[\mu} i\mathcal M_1^{\nu]\Delta} $\\ \vphantom{c}}  & \\ [3.5ex] \hline
$j\geq 1$&
$\mathcal M_2^{\mu_1\nu_1,\mu_2\nu_2}$ & $\mathcal M_2^{\bullet\Delta,\bullet\Delta}$ & $P^\mu \mathcal M_2^{\bullet\Delta,\bullet\Delta}$ & \tabincell{c}{\vphantom{c}\\$P^\mu P^\nu \mathcal M_2^{\bullet\Delta,\bullet\Delta}$\\ $\Delta^\mu\Delta^\nu \mathcal M_2^{\bullet\Delta,\bullet\Delta}$ \\ $g^{\mu\nu}\mathcal M_2^{\bullet\Delta,\bullet\Delta}$ \\ $\mathcal M_2^{\mu\bullet,\nu\bullet}$\\$\Delta^{\{\mu}\mathcal M_2^{\nu\}\bullet,\bullet\Delta}$\\$\Delta^{[\mu}\mathcal M_2^{\nu]\bullet,\bullet\Delta}$\\ $\mathcal M_2^{\mu\Delta,\nu\Delta}\,(*)$\\ \vphantom{c}}  & \\ [3.5ex] \hline
$j\geq\frac{3}{2}$&
$\mathcal M_3^{\mu_1\nu_1,\mu_2\nu_2,\mu_3\nu_3}$ & $-$ &$i\mathcal M_3^{\mu\Delta,\bullet\Delta,\bullet\Delta}$&\tabincell{c}{\vphantom{c}\\$P^{\{\mu} i\mathcal M_3^{\nu\}\Delta,\bullet\Delta,\bullet\Delta}$\\ $P^{[\mu} i\mathcal M_3^{\nu]\Delta,\bullet\Delta,\bullet\Delta}$\\ \vphantom{c}}   & \\ [3.5ex] \hline
$j\geq 2$&
$\mathcal M_4^{\mu_1\nu_1,\mu_2\nu_2,\mu_3\nu_3,\mu_4\nu_4}$ & $\mathcal M_4^{\bullet\Delta,\bullet\Delta,\bullet\Delta,\bullet\Delta}$ & $P^\mu \mathcal M_4^{\bullet\Delta,\bullet\Delta,\bullet\Delta,\bullet\Delta}$& \tabincell{c}{\vphantom{c}\\$P^\mu P^\nu \mathcal M_4^{\bullet\Delta,\bullet\Delta,\bullet\Delta,\bullet\Delta}$ \\ $\Delta^\mu\Delta^\nu \mathcal M_4^{\bullet\Delta,\bullet\Delta,\bullet\Delta,\bullet\Delta}$ \\ $g^{\mu\nu}\mathcal M_4^{\bullet\Delta,\bullet\Delta,\bullet\Delta,\bullet\Delta}$ \\ $\mathcal M_4^{\mu\bullet,\nu\bullet,\bullet\Delta,\bullet\Delta}$\\$\Delta^{\{\mu}\mathcal M_4^{\nu\}\bullet,\bullet\Delta,\bullet\Delta,\bullet\Delta}$\\$\Delta^{[\mu}\mathcal M_4^{\nu]\bullet,\bullet\Delta,\bullet\Delta,\bullet\Delta}$\\ $\mathcal M_4^{\mu\Delta,\nu\Delta,\bullet\Delta,\bullet\Delta}\,(*)$\\ \vphantom{c}}    & \\ [3.5ex] \hline\hline
\bf{Total number} &$2j+1$  & $\lfloor j\rfloor+1$ &$2j+1$& $4j+5\lfloor j\rfloor+3-\theta(j\geq 1) $ & \\ [3.5ex] \hline
\end{tabular}
\caption{Summary of the linearly independent structures appearing in the parametrisation of the matrix element of local operators, built from the covariant multipoles. In the case of a tensor operator, the term with a symbol $(*)$ built from the even multipole $\mathcal M_{2\lfloor j\rfloor}$ does not appear.}
\label{Table_summary}
\end{table}

\section{Conclusions and outlook}

In this work we have derived the general Lorentz covariant form factor (FF) parametrisations for the matrix elements of local scalar, vector, and tensor operators for massive particle states of arbitrary spin. 
We have followed two distinct and complementary approaches. The first one, which we refer to as the tensor product approach, follows the spirit of the existing literature, where  all  possible structures that build the parametrisation are explicitly derived. We recalled the known vector case and extended the treatment to the rank-$2$ tensor operator, especially relevant because of the applications to the energy-momentum tensor. We found that the counting of FFs in the tensor case depends on the spin value $j$ in a non-trivial way. 

The second approach, of central importance for our work, is based on the expansion of the operators in terms of covariant multipoles, built from the Lorentz generators in the chosen spin representation. The latter technique is especially useful because it underlines the intrinsic universal properties of the multipole expansions, independently of the particle spin and the operator type. The fundamental basis of linearly independent multipoles, constructed from symmetric and traceless products of Lorentz generators, can be used as a universal starting point for analysing the arbitrary spin matrix elements of any operator, including those that are non-local and of higher-rank. 
The specificity of the problem, such as the operator type and the spin of the particle state, dictates the actual arrangements of the multipole elements. 
Namely, the rank of the operator is responsible for selecting the allowed contractions that give rise to the various structures in front of the FFs, which are different in the scalar, vector, and tensor cases.
The value of the particle spin is, on the other hand, responsible for the truncation of the multipole series at a certain order. We find an exact
correspondence between the FF counting rules in the two different approaches, as one would expect since the total number of independent FFs cannot depend upon the bases chosen for the parametrisation.
 
As previously mentioned, a natural extension of this work would be to apply the covariant multipole expansion to the matrix elements of non-local currents. This could be achieved via the introduction of an additional vector $n^\mu$ which defines the light-front direction along which the operator is non-local~\cite{Lorce:2015lna}. Non-local currents have potential applications for many observables in QCD, including those that are expressed in terms of standard parton distributions and their generalisations (PDFs, GPDs, TMDs, etc.). This interesting follow-up is left for future work.

\section*{Acknowledgements}
The authors are grateful to M. Polyakov and K.M. Semenov-Tian-Shansky for useful discussions, and to V. Troitsky for drawing our attention to Ref.~\cite{Cheshkov1963}. The authors also acknowledge financial support by the Agence Nationale de la Recherche under the projects No. ANR-18-ERC1-0002 and ANR-16-CE31-0019.

\appendix

\section{Generalised polarisation tensors}\label{AppA}

An explicit construction of the GPTs has been proposed long ago~\cite{Auvil:1966eao,Carruthers:1966kee,Lee:1971ka,Huang:2003ym}. It consists of coupling the maximum possible spin of two lower-order GPTs. In the half-integer spin case it is convenient to adopt the Rarita-Schwinger approach~\cite{Rarita:1941mf}, i.e. to consider the following product of a Dirac spinor\footnote{For convenience, we choose to normalise the Dirac spinors as $\overline u(p,\lambda')u(p,\lambda)=1$ instead of $\overline u(p,\lambda')u(p,\lambda)=2M$.} with an integer-spin GPT
\begin{equation}\label{spintensor}
\begin{aligned}
u_{\alpha_1\cdots\alpha_n}(p,\lambda)&=\sum_{m,m'}\langle
\tfrac{1}{2}m,nm'|j\lambda\rangle\,u(p,m)\,\varepsilon_{\alpha_1\cdots\alpha_n}(p,m')\\
&=\sqrt{\frac{j+\lambda}{2j}}\,u(p,+\tfrac{1}{2})\,\varepsilon_{\alpha_1\cdots\alpha_n}(p,\lambda-\tfrac{1}{2})+\sqrt{\frac{j-\lambda}{2j}}\,u(p,-\tfrac{1}{2})\,\varepsilon_{\alpha_1\cdots\alpha_n}(p,\lambda+\tfrac{1}{2}),
\end{aligned}
\end{equation}
where $\langle j_1m_1,j_2m_2|jm\rangle$ represents the Clebsch-Gordan coefficient in the Condon-Shortley phase convention. In this way, one just needs to focus on the GPT for integer spin only. It can be constructed from the recursion formula
\begin{equation}\label{tensorprod}
\varepsilon_{\alpha_1\cdots\alpha_n}(p,\lambda)=\sum_{m,m'}\langle
1m,(n-1)m'|n\lambda\rangle\,\varepsilon_{\alpha_1}(p,m)\,\varepsilon_{\alpha_2\cdots\alpha_n}(p,m'),
\end{equation}
where $\varepsilon_{\alpha_i}(p,\lambda)$ is the standard polarisation four-vector. Iterating this formula, one finds
\begin{displaymath}
\varepsilon_{\alpha_1\cdots\alpha_n}(p,\lambda)=\sum_{\{m_i=0,\pm1\}}\left[\prod_{l=1}^n\langle1m_l,(n-l)m'_{n-l}|(n-l+1)m'_{n-l+1}\rangle\,\varepsilon_{\alpha_l}(p,m_l)\right],
\end{displaymath} 
where the sum is implicitly restricted to configurations such that $\sum_{i=1}^nm_i=\lambda$, and where $m'_l=\sum_{k=n-l+1}^n m_k$. Since the Clebsch-Gordan coefficients can be written as \cite{Varshalovich:1988ye}
\begin{equation}
\langle
1mlm_l|(l+1)(m+m_l)\rangle=\sqrt{\frac{C_2^{1+m}C_{2l}^{l+m_l}}{C_{2l+2}^{l+m_l+m+1}}},\qquad C_n^k=\left(\begin{array}{c}n\\ k\end{array}\right)\equiv\left\{\begin{array}{cl} \frac{n!}{k!\,(n-k)!},&n\geq k\geq 0\\ 0,& \textrm{otherwise}\end{array}\right.,
\end{equation}
one obtains the expression
\begin{displaymath}
\varepsilon_{\alpha_1\cdots\alpha_n}(p,\lambda)=\sum_{\{m_i=0,\pm1\}}\frac{\prod_{l=1}^n\sqrt{C_2^{1+m_l}}\,\varepsilon_{\alpha_l}(p,m_l)}{\sqrt{C_{2n}^{n+\lambda}}}.
\end{displaymath}
When $\lambda\geq 0$ this can be rewritten more conveniently as~\cite{Chung:1997jn} 
\begin{equation}\label{poltensor}
\varepsilon_{\alpha_1\cdots\alpha_n}(p,\lambda) =
\sum_{k=0}^{m/2}\frac{\sum_\mathcal{P}\left[\prod_{l=1}^k\varepsilon_{\alpha_{\mathcal{P}(l)}}(p,-1)\right]\left[\prod_{l=k+1}^{m-k}\varepsilon_{\alpha_{\mathcal{P}(l)}}(p,0)\right]\left[\prod_{l=m-k+1}^n\varepsilon_{\alpha_{\mathcal{P}(l)}}(p,+1)\right]}{2^{k-m/2}\,k!\,(m-2k)!\,(n-m+k)!\,\sqrt{C_{2n}^m}},
\end{equation}
where $\mathcal{P}$ stands for a permutation of $\{1,\cdots,n\}$ and $m=n-\lambda$. The expression for $\lambda<0$ is obtained using the relation
\begin{equation}
\varepsilon_{\alpha_1\cdots\alpha_n}(p,\lambda)=(-1)^\lambda\varepsilon^*_{\alpha_1\cdots\alpha_n}(p,-\lambda)
\end{equation}
and takes the form
\begin{equation}
\varepsilon_{\alpha_1\cdots\alpha_n}(p,\lambda) =
\sum_{k=0}^{m/2}\frac{\sum_\mathcal{P}\left[\prod_{l=1}^k\varepsilon_{\alpha_{\mathcal{P}(l)}}(p,+1)\right]\left[\prod_{l=k+1}^{m-k}\varepsilon_{\alpha_{\mathcal{P}(l)}}(p,0)\right]\left[\prod_{l=m-k+1}^n\varepsilon_{\alpha_{\mathcal{P}(l)}}(p,-1)\right]}{2^{k-m/2}\,k!\,(m-2k)!\,(n-m+k)!\,\sqrt{C_{2n}^m}}.
\end{equation}

\section{Useful identities}\label{AppB}

\subsection{General identities}

A particularly useful identity is the Schouten identity
\begin{equation}\label{Schouten}
i\epsilon^{\mu\nu\rho\sigma}g^{\tau\lambda}+i\epsilon^{\nu\rho\sigma\tau}g^{\mu\lambda}+i\epsilon^{\rho\sigma\tau\mu}g^{\nu\lambda}+i\epsilon^{\sigma\tau\mu\nu}g^{\rho\lambda}+i\epsilon^{\tau\mu\nu\rho}g^{\sigma\lambda}=0    
\end{equation}
which states that there cannot be totally antisymmetric tensors with rank larger than the spacetime dimension. Contracting this relation with an antisymmetric tensor $A_{\sigma\tau}=-A_{\tau\sigma}$ leads then to
\begin{equation}\label{ASchouten}
i\epsilon^{\mu\nu\rho}_{\phantom{\mu\nu\rho}\sigma}A^{\sigma\lambda}=\tilde A^{\nu\rho}g^{\mu\lambda}+\tilde A^{\rho\mu}g^{\nu\lambda}+\tilde A^{\mu\nu}g^{\rho\lambda},
\end{equation}
where we defined the dual antisymmetric tensor as
\begin{equation}
\tilde A^{\mu\nu}=-\frac{i}{2}\epsilon^{\mu\nu\rho\sigma}A_{\rho\sigma}.    
\end{equation}
In other words, an incomplete contraction between a Levi-Civita tensor with an antisymmetric tensor can always be rewritten in terms of the dual antisymmetric tensor (i.e. a full contraction). A typical antisymmetric tensor is $i\sigma^{\mu\nu}$ for which
\begin{equation}\label{tensorpseudoid}
i\tilde\sigma^{\mu\nu}=i\sigma^{\mu\nu}\gamma_5.
\end{equation}
One can also contract Eq.~\eqref{Schouten} with $a_\tau a_\lambda$ leading to
\begin{equation}\label{aaSchouten}
i\epsilon^{\mu\nu\rho\sigma}a^2+i\epsilon^{\nu\rho\sigma\tau}a_\tau a^\mu-i\epsilon^{\rho\sigma\mu\tau}a_\tau a^\nu+i\epsilon^{\sigma\mu\nu\tau}a_\tau a^\rho-i\epsilon^{\mu\nu\rho\tau}a_\tau a^\sigma=0.
\end{equation}

Note that $\epsilon_{\mu\nu\rho\sigma}$ and $\gamma_5=\frac{i}{4!}\epsilon_{\mu\nu\rho\sigma}\gamma^\mu\gamma^\nu\gamma^\rho\gamma^\sigma$ can be discarded from our parametrisations. Indeed, amplitudes with intrinsic positive parity necessarily involve an even number of $\epsilon_{\mu\nu\rho\sigma}$ (possibly in the form of $\gamma_5$), and the product $\epsilon_{\mu\nu\rho\sigma}\epsilon_{\alpha\beta\tau\lambda}$ can always be rewritten in terms of the metric only.
\begin{equation}
\epsilon_{\mu\nu\rho\sigma}\epsilon_{\alpha\beta\tau\lambda}=-\begin{vmatrix}
g_{\mu\alpha}&g_{\mu\beta}&g_{\mu\tau}&g_{\mu\lambda}\\
g_{\nu\alpha}&g_{\nu\beta}&g_{\nu\tau}&g_{\nu\lambda}\\
g_{\rho\alpha}&g_{\rho\beta}&g_{\rho\tau}&g_{\rho\lambda}\\
g_{\sigma\alpha}&g_{\sigma\beta}&g_{\sigma\tau}&g_{\sigma\lambda}
\end{vmatrix}.
\end{equation}

\subsection{On-shell identities}

On-shell polarisation four-vectors are orthogonal to their four-momentum argument
\begin{equation}
    p\cdot\varepsilon(p,\lambda)=0,\qquad p'\cdot\varepsilon^*(p',\lambda')=0.
\end{equation}
In terms of the symmetric variables $P=(p'+p)/2$ and $\Delta=p'-p$, these on-shell conditions read
\begin{equation}\label{onshelleps}
    P\cdot\varepsilon(p,\lambda)=\Delta\cdot\varepsilon(p,\lambda)/2,\qquad P\cdot\varepsilon^*(p',\lambda')=-\Delta\cdot\varepsilon^*(p',\lambda')/2.
\end{equation}
Since we reserved the indices $\alpha_i$ and $\alpha'_i$ to GPTs, we can write
\begin{equation}\label{PDelta}
P^{\alpha_i}\doteq\frac{\Delta^{\alpha_i}}{2},\qquad P^{\alpha'_i}\doteq-\frac{\Delta^{\alpha'_i}}{2},
\end{equation}
where $\doteq$ means on-shell equality, i.e. equality once contracted with GPTs like in~\eqref{onshelleps}. In our parametrisations, we chose to eliminate the contractions $P\cdot\varepsilon$ and $P\cdot\varepsilon^*$.

Dirac bilinears also satisfy a number of on-shell relations derived using the Dirac equation. The most famous one is the Gordon identity
\begin{equation}
\overline u(p',\lambda')\gamma^\mu u(p,\lambda)=\overline u(p',\lambda')\left[\frac{P^\mu}{M}+\frac{i\sigma^{\mu\nu}\Delta_\nu}{2M}\right]u(p,\lambda)
\end{equation}
which can be rewritten using on-shell equality as
\begin{equation}\label{Gordonid}
\gamma^\mu\doteq\frac{P^\mu}{M}+\frac{i\sigma^{\mu\nu}\Delta_\nu}{2M}.
\end{equation}
The (over-)complete set of on-shell relations has been derived in~\cite{Lorce:2017isp} and reads
\begin{align}
\mathds 1&\doteq\frac{\slashed P}{M}, &0&\doteq\slashed\Delta,\label{scalarid}\\
\gamma_5&\doteq\frac{\slashed\Delta\gamma_5}{2M}, &0&\doteq\slashed P\gamma_5,\label{pseudoscalarid}\\
\gamma^\mu&\doteq\frac{P^\mu}{M}+\frac{i\sigma^{\mu\Delta}}{2M},&0&\doteq\frac{\Delta^\mu}{2}+i\sigma^{\mu P},\label{vectorid}\\
\gamma^\mu\gamma_5&\doteq\frac{\Delta^\mu\gamma_5}{2M}+\frac{i\sigma^{\mu P}}{M},&0&\doteq P^\mu\gamma_5+\frac{i\sigma^{\mu\Delta}\gamma_5}{2},\label{pseudovectorid}\\
i\sigma^{\mu\nu}&\doteq-\frac{\Delta^{[\mu}\gamma^{\nu]}}{2M}+\frac{i\epsilon^{\mu\nu P\lambda}\gamma_\lambda\gamma_5}{M},&0&\doteq-P^{[\mu}\gamma^{\nu]}+\frac{i\epsilon^{\mu\nu\Delta\lambda}\gamma_\lambda\gamma_5}{2},\label{tensorid}\\
i\sigma^{\mu\nu}\gamma_5&\doteq-\frac{P^{[\mu}\gamma^{\nu]}\gamma_5}{M}+\frac{i\epsilon^{\mu\nu\Delta\lambda}\gamma_\lambda}{2M},&0&\doteq-\frac{\Delta^{[\mu}\gamma^{\nu]}\gamma_5}{2}+i\epsilon^{\mu\nu P\lambda}\gamma_\lambda,
\end{align}
where a contraction with a four-vector is denoted by replacing a Lorentz index by the four-vector, like e.g. $i\sigma^{\mu\Delta}\equiv i\sigma^{\mu\nu}\Delta_\nu$. In our parametrisations, we chose to parametrise the parity-even sector in terms of $\mathds 1$ and $i\sigma^{\mu\nu}$ and eliminated $i\sigma^{\mu P}$ using~\eqref{vectorid}. Combining Eqs.~\eqref{vectorid} and~\eqref{tensorid} we find
\begin{equation}
P^2 i\sigma^{\mu\Delta}\doteq\frac{\Delta^2}{2}\,P^\mu-Mi\epsilon^{\mu P\Delta\lambda}\gamma_\lambda\gamma_5.
\end{equation}
Combining Eqs.~\eqref{ASchouten}, \eqref{tensorpseudoid}, and~\eqref{pseudovectorid} we also get
\begin{equation}\label{epsgamma5}
    -2i\epsilon^{\mu\nu\rho P}\gamma_5\doteq \Delta^\mu i\sigma^{\nu\rho}+\Delta^\nu i\sigma^{\rho\mu}+\Delta^\rho i\sigma^{\mu\nu}.
\end{equation}
Together with Eq.~\eqref{aaSchouten}, this allows us to eliminate the product $\epsilon^{\mu\nu\rho\sigma} \gamma_5$ from our parametrisations.

Since Rarita-Schwinger spinors satisfy the constraint
\begin{equation}\label{RSconstraint}
\gamma^{\alpha_i} u_{\alpha_1\cdots\alpha_n}(p,\lambda)=0  ,\qquad  \overline u_{\alpha'_1\cdots\alpha'_n}(p',\lambda')\gamma^{\alpha'_i}=0,\qquad i\in\{1,\cdots,n\}
\end{equation}
the structures $i\sigma^{\mu\nu}$ carrying an $\alpha_i$ or $\alpha'_i$ index can also be eliminated owing to
\begin{equation}\label{sigmaelim}
i\sigma^{\alpha'\mu}\doteq g^{\alpha'\mu},\qquad i\sigma^{\nu\alpha}\doteq g^{\nu\alpha}.
\end{equation}
We can derive a number of interesting relations. Starting from the product of four Dirac matrices
\begin{equation}
\begin{aligned}
\gamma^\rho\gamma^\mu\gamma^\nu\gamma^\sigma&=g^{\rho\mu}g^{\nu\sigma}-g^{\rho\nu}g^{\mu\sigma}+g^{\rho\sigma}g^{\mu\nu}+i\epsilon^{\rho\mu\nu\sigma}\gamma_5\\
&-g^{\rho\mu}i\sigma^{\nu\sigma}+g^{\rho\nu}i\sigma^{\mu\sigma}-g^{\rho\sigma}i\sigma^{\mu\nu}\\
&-i\sigma^{\rho\mu}g^{\nu\sigma}+i\sigma^{\rho\nu}g^{\mu\sigma}-i\sigma^{\rho\sigma}g^{\mu\nu},
\end{aligned}
\end{equation}
we find using Eqs.~\eqref{RSconstraint} and~\eqref{sigmaelim}
\begin{subequations}
\begin{align}
i\epsilon^{\rho\mu\nu\alpha}\gamma_5&\doteq g^{\alpha\rho}i\sigma^{\mu\nu}+g^{\alpha\mu}i\sigma^{\nu\rho}+g^{\alpha\nu}i\sigma^{\rho\mu},\\
i\epsilon^{\alpha'\mu\nu\sigma}\gamma_5&\doteq g^{\alpha'\sigma}i\sigma^{\mu\nu}+g^{\alpha'\mu}i\sigma^{\nu\sigma}+g^{\alpha'\nu}i\sigma^{\sigma\mu},\\
i\epsilon^{\alpha'\mu\nu\alpha}\gamma_5&\doteq g^{\alpha'\alpha}i\sigma^{\mu\nu}+g^{\alpha'\mu}g^{\nu\alpha}-g^{\alpha'\nu}g^{\mu\alpha}.\label{niceid1}
\end{align}
\end{subequations}
Contracting the last relation with $P$ and $\Delta$ we get using Eq.~\eqref{vectorid}
\begin{subequations}\label{eps4id}
\begin{align}
i\epsilon^{\alpha' P\nu\alpha}\gamma_5&\doteq\frac{1}{2}\,g^{\alpha'\alpha}\Delta^\nu+P^{[\alpha'}g^{\alpha]\nu},\\    
i\epsilon^{\alpha'\Delta\nu\alpha}\gamma_5&\doteq-g^{\alpha'\alpha}i\sigma^{\nu\Delta}+\Delta^{[\alpha'}g^{\alpha]\nu},\\
i\epsilon^{\alpha' P\Delta\alpha}\gamma_5&\doteq\frac{1}{2}\,g^{\alpha'\alpha}\Delta^2+P^{[\alpha'}\Delta^{\alpha]}.
\end{align}
\end{subequations}
From the product of three Dirac matrices
\begin{align}
\gamma^\rho\gamma^\mu\gamma^\sigma&=g^{\rho\mu}\gamma^\sigma-g^{\rho\sigma}\gamma^\mu+g^{\mu\sigma}\gamma^\rho-i\epsilon^{\rho\mu\sigma\lambda}\gamma_\lambda\gamma_5,\\
\gamma^\rho\gamma^\mu\gamma^\sigma\gamma_5&=g^{\rho\mu}\gamma^\sigma\gamma_5-g^{\rho\sigma}\gamma^\mu\gamma_5+g^{\mu\sigma}\gamma^\rho\gamma_5-i\epsilon^{\rho\mu\sigma\lambda}\gamma_\lambda,
\end{align}
we find using Eq.~\eqref{RSconstraint}
\begin{subequations}\label{3gamma}
\begin{align}
i\epsilon^{\rho\mu\alpha\lambda}\gamma_\lambda\gamma_5&\doteq g^{\mu\alpha}\gamma^\rho-g^{\rho\alpha}\gamma^\mu,&i\epsilon^{\rho\mu\alpha\lambda}\gamma_\lambda&\doteq g^{\mu\alpha}\gamma^\rho\gamma_5-g^{\rho\alpha}\gamma^\mu\gamma_5,\\
i\epsilon^{\alpha'\mu\sigma\lambda}\gamma_\lambda\gamma_5&\doteq g^{\alpha'\mu}\gamma^\sigma-g^{\alpha'\sigma}\gamma^\mu,&i\epsilon^{\alpha'\mu\sigma\lambda}\gamma_\lambda&\doteq g^{\alpha'\mu}\gamma^\sigma\gamma_5-g^{\alpha'\sigma}\gamma^\mu\gamma_5,\\
i\epsilon^{\alpha'\mu\alpha\lambda}\gamma_\lambda\gamma_5&\doteq -g^{\alpha'\alpha}\gamma^\mu,&i\epsilon^{\alpha'\mu\alpha\lambda}\gamma_\lambda&\doteq -g^{\alpha'\alpha}\gamma^\mu\gamma_5.
\end{align}
\end{subequations}
Combining these with Eqs.~\eqref{Schouten} and~\eqref{pseudoscalarid}, we find
\begin{equation}\label{niceid2}
i\epsilon^{\alpha'\mu\nu\alpha}\gamma_5\doteq\frac{1}{2M}\left[(g^{\mu\alpha}\gamma^\nu-g^{\nu\alpha}\gamma^\mu)\,\Delta^{\alpha'}+(g^{\mu\alpha'}\gamma^\nu-g^{\nu\alpha'}\gamma^\mu)\,\Delta^\alpha+g^{\alpha'\alpha}(\gamma^\mu\Delta^\nu-\gamma^\nu\Delta^\mu)\right].
\end{equation}
This relation is quite remarkable since it allows one to replace a structure antisymmetric under $\alpha'\leftrightarrow\alpha$ by a symmetric one, which can of course only be true on shell. Contracting Eqs.~\eqref{3gamma} with $P$ and $\Delta$, and using Eq.~\eqref{scalarid} and~\eqref{pseudoscalarid}, we get
\begin{subequations}\label{gamma3}
\begin{align}
i\epsilon^{\rho P\alpha\lambda}\gamma_\lambda\gamma_5&\doteq P^\alpha\gamma^\rho-Mg^{\rho\alpha},&i\epsilon^{\rho P\alpha\lambda}\gamma_\lambda&\doteq P^\alpha\gamma^\rho\gamma_5,\\
i\epsilon^{\rho \Delta\alpha\lambda}\gamma_\lambda\gamma_5&\doteq\Delta^\alpha\gamma^\rho,&i\epsilon^{\rho \Delta\alpha\lambda}\gamma_\lambda&\doteq \Delta^\alpha\gamma^\rho\gamma_5-2Mg^{\rho\alpha}\gamma_5,\\
i\epsilon^{P\Delta\alpha\lambda}\gamma_\lambda\gamma_5&\doteq M\Delta^\alpha,&i\epsilon^{P\Delta\alpha\lambda}\gamma_\lambda&\doteq -2MP^\alpha\gamma_5,\\
i\epsilon^{\alpha' P\sigma\lambda}\gamma_\lambda\gamma_5&\doteq P^{\alpha'}\gamma^\sigma-Mg^{\alpha'\sigma},&i\epsilon^{\alpha' P\sigma\lambda}\gamma_\lambda&\doteq P^{\alpha'}\gamma^\sigma\gamma_5,\\
i\epsilon^{\alpha' \Delta\sigma\lambda}\gamma_\lambda\gamma_5&\doteq\Delta^{\alpha'}\gamma^\sigma,&i\epsilon^{\alpha' \Delta\sigma\lambda}\gamma_\lambda&\doteq \Delta^{\alpha'}\gamma^\sigma\gamma_5-2Mg^{\alpha'\sigma}\gamma_5,\\
i\epsilon^{\alpha' P\Delta\lambda}\gamma_\lambda\gamma_5&\doteq -M\Delta^{\alpha'},&i\epsilon^{\alpha' P\Delta \lambda}\gamma_\lambda&\doteq 2MP^{\alpha'}\gamma_5,\\
i\epsilon^{\alpha' P\alpha\lambda}\gamma_\lambda\gamma_5&\doteq -Mg^{\alpha'\alpha},&i\epsilon^{\alpha' P\alpha\lambda}\gamma_\lambda&\doteq 0,\label{niceid3}\\
i\epsilon^{\alpha' \Delta\alpha\lambda}\gamma_\lambda\gamma_5&\doteq 0,&i\epsilon^{\alpha' \Delta\alpha\lambda}\gamma_\lambda&\doteq -2Mg^{\alpha'\alpha}.
\end{align}
\end{subequations}
If we contract Eqs.~\eqref{niceid1} and~\eqref{niceid2} with $\Delta_\nu$ and use the Gordon identity~\eqref{Gordonid}, we obtain the non-trivial relation quoted without proof in~\cite{Nozawa:1990gt}
\begin{equation}\label{Nozawaid}
\Delta^{\alpha'}g^{\mu\alpha}-\Delta^{\alpha}g^{\mu\alpha'}\doteq 2M\left(1-\frac{\Delta^2}{4M^2}\right)g^{\alpha'\alpha}\gamma^\mu-2g^{\alpha'\alpha}P^\mu+\frac{1}{M}\,\Delta^{\alpha'}\Delta^\alpha\gamma^\mu.
\end{equation}
Our derivation shows in particular that this on-shell relation holds not only for spin-$3/2$, but also for all higher-spin Rarita-Schwinger spinors. 
\newline 

Let us now consider a single contraction between two Levi-Civita tensors
\begin{equation}\label{epseps}
\begin{aligned}
-i\epsilon^{\rho\sigma\tau}_{\phantom{\rho\sigma\tau}\lambda}i\epsilon^{\lambda\alpha\beta\mu}&=g^{\rho\alpha}g^{\sigma\beta}g^{\tau\mu}+g^{\rho\beta}g^{\sigma\mu}g^{\tau\alpha}+g^{\rho\mu}g^{\sigma\alpha}g^{\tau\beta}\\
&-g^{\rho\beta}g^{\sigma\alpha}g^{\tau\mu}-g^{\rho\alpha}g^{\sigma\mu}g^{\tau\beta}-g^{\rho\mu}g^{\sigma\beta}g^{\tau\alpha}.
\end{aligned}
\end{equation}
Using Schouten identity~\eqref{Schouten} and Eq.~\eqref{scalarid}, we have
\begin{equation}
    -i\epsilon^{P\Delta\mu\nu}\gamma^\lambda+M i\epsilon^{\lambda\Delta\mu\nu}\doteq  i\epsilon^{\lambda P\Delta[\mu}\gamma^{\nu]}.
\end{equation}
Contracting this with $-i\epsilon^{\alpha'\alpha P}_{\phantom{\alpha'\alpha P}\lambda}$ gives then
\begin{equation}
M\Delta^{[\alpha'}g^{\alpha][\mu}P^{\nu]}\doteq P^{[\alpha'}\Delta^{\alpha]}P^{[\mu}\gamma^{\nu]}+P^2\Delta^{[\alpha'}g^{\alpha][\mu}\gamma^{\nu]},
\end{equation}
where we used Eqs.~\eqref{niceid3} and~\eqref{epseps}. Finally, thanks to Gordon identity~\eqref{Gordonid} and Eq.~\eqref{PDelta} we obtain
\begin{equation}
\frac{\Delta^2}{2}\,\Delta^{[\alpha'}g^{\alpha][\mu}P^{\nu]}\doteq -\Delta^{\alpha'}\Delta^\alpha P^{[\mu}i\sigma^{\nu]\Delta}+P^2\Delta^{[\alpha'}g^{\alpha][\mu}i\sigma^{\nu]\Delta}.
\end{equation}

Another non-trivial relation can be found as follows. Using the Schouten identity~\eqref{Schouten} we can write
\begin{equation}
-g^{[\alpha'\{\mu}i\epsilon^{\alpha]\nu\}P\Delta}\gamma_5=2g^{\mu\nu}i\epsilon^{P\Delta\alpha'\alpha}\gamma_5-P^{\{\mu}i\epsilon^{\nu\}\Delta\alpha'\alpha}\gamma_5+\Delta^{\{\mu}i\epsilon^{\nu\}P\alpha'\alpha}\gamma_5
\end{equation}
which leads to
\begin{equation}
-g^{[\alpha'\{\mu}i\epsilon^{\alpha]\nu\}P\Delta}\gamma_5\doteq \Delta^2g^{\mu\nu}g^{\alpha'\alpha}-2g^{\mu\nu}\Delta^{\alpha'}\Delta^\alpha-g^{\alpha'\alpha}P^{\{\mu}i\sigma^{\nu\}\Delta}+\Delta^{[\alpha'}g^{\alpha]\{\mu}P^{\nu\}}-g^{\alpha'\alpha}\Delta^\mu\Delta^\nu+\frac{1}{2}\,\Delta^{\{\alpha'}g^{\alpha\}\{\mu}\Delta^{\nu\}}   
\end{equation}
using Eqs.~\eqref{eps4id}. Alternatively, using Eq.~\eqref{scalarid} and again the Schouten identity gives
\begin{equation}
-g^{[\alpha'\{\mu}i\epsilon^{\alpha]\nu\}P\Delta}\gamma_5\doteq\frac{1}{2M}\,g^{[\alpha'\{\mu}\left(\Delta^{\alpha]}i\epsilon^{\nu\}P\Delta\lambda}-\Delta^{\nu\}}i\epsilon^{\alpha]P\Delta\lambda}-\Delta^2i\epsilon^{\alpha]\nu\}P\lambda}\right)\gamma_\lambda\gamma_5    
\end{equation}
which leads to
\begin{equation}
-g^{[\alpha'\{\mu}i\epsilon^{\alpha]\nu\}P\Delta}\gamma_5\doteq \frac{1}{2}\,\Delta^{[\alpha'}g^{\alpha]\{\mu}i\sigma^{\nu\}\Delta}-\frac{1}{2}\,\Delta^{\{\alpha'}g^{\alpha\}\{\mu}\Delta^{\nu\}}+\Delta^2g^{\alpha'\{\mu}g^{\nu\}\alpha}   
\end{equation}
using Eqs.~\eqref{gamma3} and~\eqref{PDelta}.

\bibliography{main}

\end{document}